\documentclass[11pt]{article}
\usepackage{feynmp}
\usepackage{slashed}
\usepackage{graphicx}
\usepackage{epstopdf}
\usepackage{subfigure}
\usepackage{amsmath}
\usepackage{amsfonts}
\usepackage{amssymb}
\usepackage{pifont}
\usepackage{tikz}
\usepackage{color}
\usepackage{mathrsfs}
\usepackage[colorlinks=true, linkcolor=blue, citecolor=blue, urlcolor=blue]{hyperref}

\numberwithin{equation}{section}

\title{\textbf{A CFT Perspective on\\[2.5mm]
Gravitational Dressing and Bulk Locality\\[4mm]}}

\author{\text{Aitor Lewkowycz,}\thanks{\href{mailto:lewkow@gmail.com}{\protect\path{lewkow@gmail.com}}}\vspace{0.2cm}
 \text{  Gustavo J. Turiaci}\thanks{\href{mailto:turiaci@princeton.edu}{\protect\path{turiaci@princeton.edu}}}\vspace{0.2cm}
\text{ and Herman Verlinde}\thanks{\href{mailto:verlinde@princeton.edu}{\protect\path{verlinde@princeton.edu}}}\vspace{0.2cm}\\[2mm]
\small\textit{Physics Department}\\ \small\textit{Princeton University,}\\\small \textit{NJ 08544, USA}
\vspace{0.5cm}
\\
\small \textit{${}^\ddagger$Princeton Center for Theoretical Science,}\\
\small \textit{Princeton University,}\\
 \small\textit{NJ 08544, USA}}
\date{}
\begin{document}
\maketitle
\thispagestyle{empty}
\begin{abstract}

We revisit the construction of local bulk operators in AdS/CFT with special focus on  gravitational dressing and its consequences for bulk locality.  Specializing to 2+1-dimensions, we investigate these issues via the proposed identification between bulk operators and cross-cap boundary states. We obtain explicit expressions for correlation functions of bulk fields with boundary stress tensor insertions, and find that they are free of non-local branch cuts but do have non-local poles. 
We recover the HKLL recipe for restoring bulk locality for interacting fields 
as the outcome of a natural CFT crossing condition. We show that, in a suitable gauge, the cross-cap states solve the bulk wave equation for general background geometries, and satisfy a conformal Ward identity analogous to a soft graviton theorem. Virasoro symmetry, the large $N$ conformal bootstrap and the uniformization theorem all play a key role in our derivations.

\end{abstract}
\newpage
\tableofcontents
\thispagestyle{empty}
\newpage
\setcounter{page}{1}
\def\ttiny{\scriptsize}

 \setlength{\textwidth}{460pt}
\setlength{\topmargin}{-1.2cm}
\setlength{\textheight}{630pt}
\setlength{\oddsidemargin}{10pt}
\linespread{1.25}
\def\tildez{{z'}}
\newcommand{\beq}{\begin{equation}}
\newcommand{\eeq}{\end{equation}}
\newcommand{\nn}{\nonumber\\} 
\newcommand{\bea}{\begin{eqnarray}}
\newcommand{\ea}{\end{eqnarray}}
\newcommand{\barr}{\begin{array}}
\newcommand{\earr}{\end{array}}
\newcommand{\lb}{{\langle}}
\newcommand{\rb}{{\rangle}}

\def\d{{\partial}}
\def\n{{\bf \widehat n}}
\def\k{{\bf k}}
\newcommand{\na}{\mbox{\boldmath$\nabla$}}
\newcommand{\om}{\omega}
\newcommand{\calo}{{\cal O}}
\newcommand{\calL}{{\cal L}}
\newcommand{\vphi}{{\varphi}}
\newcommand{\zb}{\overline{z}}
\def\changemargin#1#2{\list{}{\rightmargin#2\leftmargin#1}\item[]}
\let\endchangemargin=\endlist 
\def\spc{\hspace{1pt}}
\def\nmpc{\hspace{-.5pt}}
\def\nspc{\hspace{-1pt}}
\def\smpc{\hspace{.5pt}}
\def\li{\bigl |}
\def\ra{\bigr \rangle}
\def\la{\bigl\langle}

\def\bea{\begin{eqnarray}}
\def\eea{\end{eqnarray}}
\def\is{\! & \! = \! & \!}

\def\li{\bigl|\spc}
\def\ri{\bigr |\spc}
\def\zs{z}
\def\zb{\bar{z}}

\def\calO{{b}}
\def\be{\begin{equation}}
\def\ee{\end{equation}}
\def\calO{{\Phi}}

\def\uuU{\mbox{\small$W$}}
\def\vvV{\mbox{\small$\,\overline{\!W\!}\,$}}
\def\xX{\mbox{\small$\nspc X\nspc\smpc$}}
\def\UUU{{{\!}_{{}^{W\nspc \smpc W\nspc}}}}
\def\VVV{{{\!}_{{}^{\,\overline{\!W\!}\,\nspc \smpc\,\overline{\!W\!}\,\nspc}}}}
\def\UU{{{\!}_{{}^{Z\nspc \smpc Z\nspc}}\spc }}
\def\VV{{{\!}_{{}^{\bar{Z}\nspc \smpc \bar{Z}\nspc}}}}
\def\lL{\mbox{\small $L$}}
\def\rR{\mbox{\small $R$}}
\def\uU{\mbox{\small$Z$}}
\def\tT{\mbox{\small$T$}}
\def\vV{\mbox{\small$\bar{Z}$}}
\def\vVi{\mbox{\small$\zb_i$}}
\def\VVi{{{\!}_{{}^{Z\nspc \smpc \zs_i\nspc}}}}
\def\ssca{\mbox{\scriptsize \sc a}}

\def\R{{\mathbb R}}
\def\P{{\mathbb P}}
\def\Z{{\mathbb Z}}
\def\RP{{\mathbb {RP}}}
\def\H{{\mathbb H}}
\def\C{{\mathbb C}}

\def\mmu{{\rm a} }
\def\nnu{{ \rm b}}
\def\llambda{{\rm p}}
\def\rrho{{\rm q}} 
\def\xxi{\omega}
   \def\CCC{\mbox{\small $C$}}
   \def\PPhi{\mbox{\small $\Phi$}}
   \def\PPsi{\mbox{\small $\Psi$}}

\section{Introduction}\label{SecIntro}

\addtolength{\abovedisplayskip}{.6mm}
\addtolength{\belowdisplayskip}{.6mm}
\addtolength{\baselineskip}{.3mm}
\addtolength{\parskip}{.6mm}

In this paper we will revisit the construction of bulk operators in AdS/CFT \cite{adscft, banks}, with special focus on gravitational dressing and its consequences for locality. In particular, we will investigate to what extent these issues are incorporated in the recently proposed identification of bulk operators with cross-cap boundary states in the CFT \cite{HV, Takaya1, NO1, NO2, Takaya2}. We specialize our discussion to AdS${}_3$/CFT${}_2$.

Most investigations of bulk physics in AdS work at leading order in large $N$. In this limit, it is  reasonable to assume that a standard semi-classical description applies, and that bulk operators should look like free local quantum fields propagating on a fixed background geometry \cite{banks, Hamilton:2006az, HKLLint, KL15}. 
The infinite $N$ limit, however, is bound to be somewhat misleading. The holographic mapping relies essentially on the fact that the bulk theory is gravitational and therefore inherently non-local. Localized observables must make reference to delocalized geometric data, such as the distance to some fixed reference point or asymptotic boundary. In a quantum gravitational theory, the distance between two points defines a non-local operator.

The non-locality of the bulk theory is more than just a technical nuisance. It is responsible for many of the apparent contradictions that plague the formalism of QFT in curved space, and is likely to be an essential component of any future microscopic explanation of holography.
It also provides an immediate reality check for the bulk reconstruction program. Rather than trying to enforce exact compliance with bulk locality, it is necessary to come to grips with the fact that bulk observables are necessarily non-local, and instead use the analytic and geometric structure of CFT as a guide for how to deal with this non-locality in a natural and practical way.

\def\yY{\mbox{\small $Y$}}
\def\tTb{\spc\overline{\nspc\tT\nspc}\spc}
\def\zZ{\mbox{\small $Z$}}
\def\tT{\mbox{$\Omega$}}
\def\scx{{\mbox{\scriptsize \sc x}}}
\def\xxX{\mbox{\small $\cal X$}}

A useful analogy for gravitational dressing is the use of Wilson lines, or generalizations thereof, to define physical operators in gauge theory. Equivalently, one could fix the gauge so that local matter fields become gauge invariant. However, their commutation relations would still be non-local. The gravitational story is quite analogous \cite{Donnelly:2015hta,marolfmintun}. In this case the dressing enforces diffeomorphism invariance and 
creates the gravitational field associated with a local operator. It also serves as an anchor for the bulk point relative to one or more reference points on the boundary. In addition, it should provide an dynamical adjustment mechanism that ensures that the bulk equation of motion remains satisfied in non-trivial background geometries. 

As in gauge theory, the gravitational dressing operation is not unique. So let us list some principles that can guide us toward a natural and  practical choice.

A set of reasonable requirements on the gravitational dressing prescription are that it should~be:

\vspace{-3mm}

\begin{enumerate}
\item{{\it Geometric}. Gravitational dressing provides a geometric anchoring for the bulk point relative to the boundary. The dressing operator should  include a projection that fixes the geodesic distance between the bulk point to a collection of boundary points.   }
\item{{\it Intrinsic}. The bulk geometry and gravitational dynamics emerge from the CFT. A non-perturbative definition of bulk operators should preferably be given in terms of intrinsic CFT data, and not presume the existence of a semi-classical bulk geometry. }
\item{{\it Minimal}.  A gravitational dressing is minimal if a correlation function of $N$ scalar bulk operators $\Phi(X_i)$ depends on at most $N(d+1)$ independent parameters, i.e. the positions~$X_i$.  Ideally, there should exist a gauge choice in which the dressing operator is invisible.}
\item{{\it Analytic}. Gravitationally dressed bulk operators are non-local, both in the bulk and in the CFT. To preserve some form of locality, we require that correlation functions between bulk and boundary operators are analytic and do not contain any unphysical branch cuts.}
\item{{\it Gauge invariant.} The stress energy tensor of the CFT generates coordinate transformations on the boundary of AdS. Diffeomorphism invariance of the bulk theory
implies that bulk correlation functions must satisfy Ward identities, or `soft graviton theorems', that generalize the conformal Ward identities of the CFT.}
\item{{\it Gauge redundant.} Physical arguments, such as the AdS/Rindler wedge construction and the apparent similarity between holographic reconstruction and quantum error correction \cite{error}, indicate that the bulk observables are not unique and should have a built-in gauge redundancy. This will help avoid or conceal the physical effect of non-local commutators.}

\end{enumerate}
\def\Phizz{{\Phi^{\mbox{\ttiny (0)}}}}
\def\Phiz{\Phi}
 
\def\etat{\eta}

In the following sections we will elaborate each of these requirements, and along the way, test to which extent they are realized by the proposed identification of bulk operators with cross-cap boundary states  \cite{HV, Takaya1, NO1}.  We mostly restrict ourselves to the special case of AdS${}_3$/CFT${}_2$. We will argue that in this case the gravitational dressing necessitates replacing the free HKLL operators (which define global cross-cap operators, that satisfy a finite number of invariance conditions \cite{NO1}) with Virasoro cross-cap operators  (that satisfy an infinite number of invariance conditions \cite{HV,Ishibashi}).  We explicitly verify that this replacement helps eliminate  unphysical branch cut singularities in graviton amplitudes. We also describe how, for holographic CFTs, approximate bulk locality emerges as a dynamical consequence of a CFT crossing relation. Moreover, we will find that the amplitudes of Virasoro cross-cap states satisfy a natural set of conformal Ward identities, that can be viewed as the AdS${}_3$ analogue of soft-graviton theorems, and a form of background independence. 
 A key geometric ingredient that underlies most of these result is the uniformization theorem. 

\section{A geometric definition of bulk operators} \label{SecBulkOp}

 We introduce two sets coordinates on AdS${}_{d+1}$
\bea
\quad \qquad \xX^{\smpc \ssca} \! \is \!  (y, x^{\mu}) \qquad \qquad \mbox{Poincar\'e}\nonumber \\[1mm]
\qquad \quad \xX \is  (\rho, t, x^\mmu) \qquad \qquad {\rm global}\nonumber
\eea
with $x^\mu$  and $(t, x^\mmu)$ coordinates along the AdS boundary ${\mathbb R} \times S^{d-1}$.

\begin{figure}[t]
\begin{center}
\includegraphics[scale=0.25]{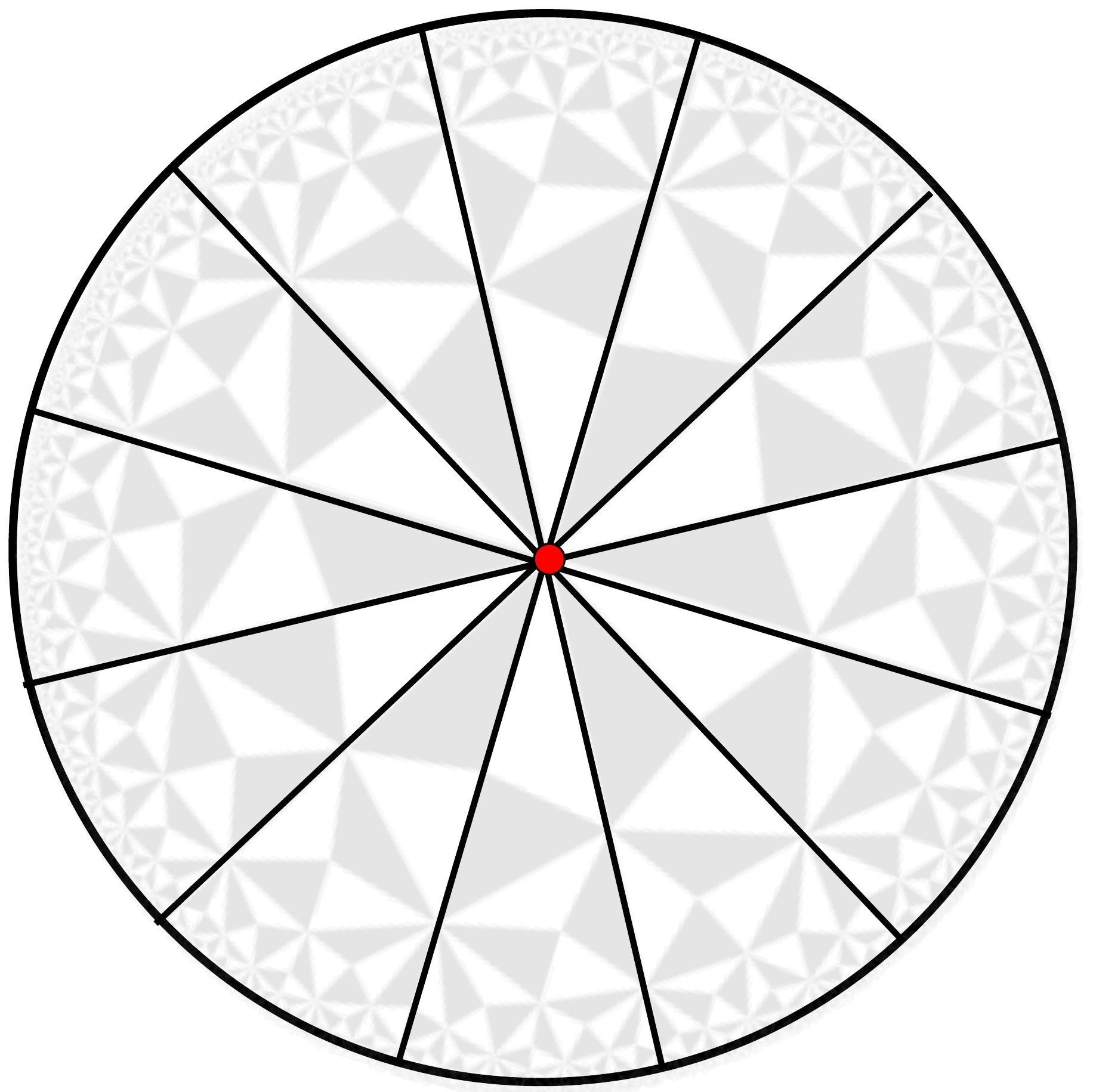}~~~~~~~~~~~~~~~~~~~~~~
\includegraphics[scale=0.25]{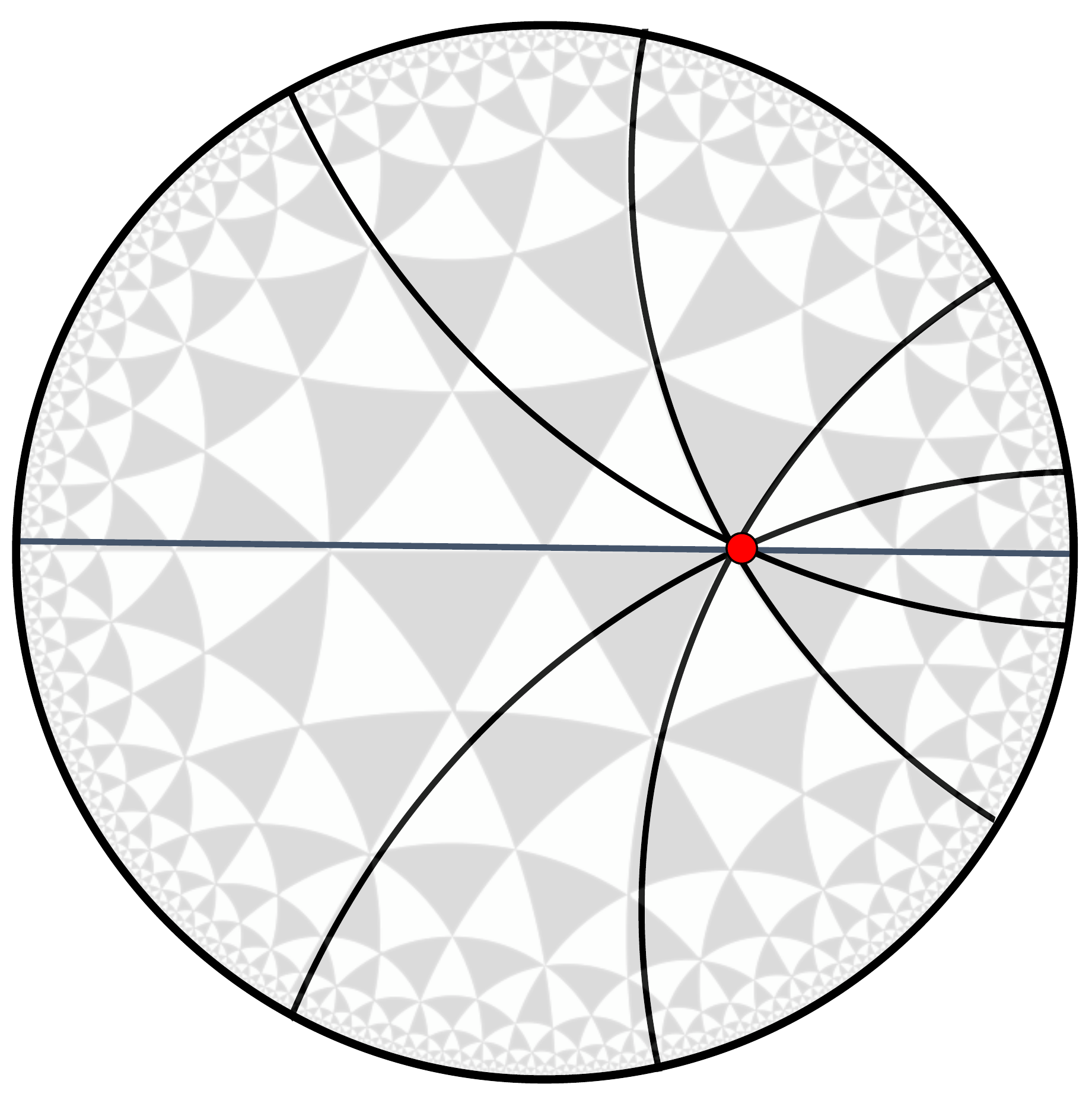}
\caption{\addtolength{\baselineskip}{-.5mm}A bulk point $X$ in AdS lies at the intersection of a continuous family of geodesics, and thereby induces an antipodal pairing  between boundary points.
 }
\label{antipodal} 
\end{center}
\vspace{-3mm}
\end{figure}

A bulk point in AdS lies at the intersection of a continuous family of geodesics. Since each geodesic starts and ends at the boundary, 
a bulk point thus induces a pairing  between boundary points, depicted in figure (\ref{antipodal}). 
When the bulk point is located at the center of AdS, the associated pairing is antipodal. We can move the point to some general location $X$ away from the center by acting with a isometry transformation which in the coordinate system described above looks like 
\bea\label{Gex}
\quad \qquad g(X) = y^{D} e^{i P_\mu  x^\mu}\, \in\; SO(d+1,2)  \qquad \qquad \mbox{Poincar\'e}\nonumber \\[1mm]
\qquad \quad  g(X) = e^{-itH} e^{\rho (P_a-K_a) x^a}\, \in\; SO(d+1,2) \qquad \qquad {\rm global}\nonumber
\eea
with $(H,M_{\mmu\nnu}, P_\mmu, K_\mmu)$  the global conformal symmetry generators on the AdS boundary $\mathbb{R}\times S^{d-1}$. $D$ is the Poincar\'e patch boundary dilatation generator and $P_\mu$ the translation generators. Both descriptions will be convenient  in the following and  we will use them interchangeably. 

The characterization of a bulk point as the common intersection of a family of geodesics is also natural from the point of view of radon transform and the kinematic space construction of bulk operators \cite{radon} \cite{kinematic}.

\subsection{Bulk operators as global cross-caps} \label{SecGlobalXcap}

The mapping that interchanges the endpoints points of all geodesics through $X$ defines an orientation reversing diffeomorphism of the boundary. For a bulk point located at $X= (y, x^\mu)$ pure AdS, this diffeomorphism takes the form of a global conformal transformation
\bea
\label{globalmap}
\tildez^\mu - x^{\smpc \mu} \, \is \, - y^2\; \frac{z^\mu - x^{\mu}}{(z-x)^2}.
\eea
We thus obtain a canonical identification between bulk points in AdS and orientation reversing global conformal transformations. Picking a bulk point breaks the 
$SO(d,2)$ group of global conformal transformations to the subgroup $SO(d-1,2)$ of transformations that commute with the  orientation reversing global conformal transformation  (\ref{globalmap}).

Bulk operators are in one-to-one correspondence with CFT operators via  the asymptotic condition $\calO(\xX)\, \to \, {y}^{2h} \, {\cal O}(x)$ for $y\to 0$.
To leading order in $1/N$ and in the AdS vacuum, the bulk field ${\Phi}^{\mbox{\ttiny (0)}}$ satisfies the free  wave equation
\bea
\label{free}
\bigl(\square + m_h^2 \bigr)\Phi^{\mbox{\ttiny (0)}}(X) = 0
\eea
with $\square$ is the wave operator in AdS${}_{d+1}$
and $m_h^2 = 2h(d-2h)$.
At sub-leading order in $1/N$, this equation of motion receives $1/N$ corrections, both due to coupling to dynamical gravity and due to self-interactions. These corrections will be the main focus of this paper.

An efficient construction of the leading order operator $\Phi^{\mbox{\ttiny (0)}}$ was introduced by Nakayama and Ooguri in \cite{NO1}. Consider the global cross-cap state $|{\Phi}^{\mbox{\ttiny (0)}}\rangle \equiv | \spc \Phizz(0)\rangle$, made from global descendents of the primary state $|{\cal O}\rangle$ of conformal dimension $h$, solving 
\bea
\label{fixedo}
M_{\mmu\nnu} \li \spc \Phizz\ra\! \is\! 0\quad \\[2mm]
\label{fixedt}
(P_\mmu + K_\mmu) \li \spc \Phizz\ra \!\is\! 0.
\eea
These equations impose the condition that $| \spc \Phizz\rangle$ is invariant under all global isometries that keep the center of AdS fixed, thus supporting its interpretation as a state  created by a local  bulk operator situated at the center. By acting with the global AdS isometries, we construct states $| \spc {\Phi}^{\mbox{\ttiny (0)}}(\xX)\rangle$ localized at any other point $\xX$
\bea
\label{gtrans}
\li \Phi^{\mbox{\ttiny \sc (0)}}(X) \ra \is g(X) 
\li \Phizz\ra
\eea
with $g(X)$ the global transformation given in equation (\ref{Gex}). The resulting operator ${\Phi}^{\mbox{\ttiny (0)}}(\xX)$ solves the free field wave equation (\ref{free}) on the unperturbed AdS geometry and satisfies the GKPW asymptotic boundary condition. 
This geometric definition of bulk operators is equivalent to the free HKLL formula $
\Phi^{\mbox{\ttiny \sc (0)}}(X) = \int\! dx \, K(X,x)\, {\cal O}(x)$
where $K(X,x)$ is the HKLL smearing function that solves the bulk equation of motion \cite{Hamilton:2006az}.

The leading order bulk operator (\ref{fixedo})-(\ref{fixedt}) can be recognized as a global cross-cap state: it imposes the antipodal identification $z = \tildez$, with $\tildez$ given in equation (\ref{globalmap}), and projects on the Hilbert space sector spanned by the global descendents of the primary operator ${\cal O}$. 
This characterization has several attractive features, and an equal number of short-comings.
It gives a nice geometric description, but it only works for pure AdS.  Moreover, the projection onto a given representation of the global conformal group is a rather artificial non-local operation that presumes a preferred global coordinate system. It is then not surprising that, as soon as interactions are included,  the leading order definition needs to be modified.

\subsection{Bulk interactions and locality I} \label{SecIntLoc1}

A first non-trivial test of locality is the three point function between a single free HKLL bulk operator and two boundary operators
\bea
\label{threept}
\la\spc \spc {\cal O}_{i}(x_1) {\cal O}_{j}(x_2)\spc\Phi_k^{\mbox{\ttiny $(0)$}}(0)\spc \ra \is \frac{(1+x_1^2)^{h_i-h_j}(1+x^2_2)^{h_j-h_i\!\!\!\!\!}}{(x_1 - x_2)^{2(h_i +h_j)}\!\!}\; \; {\rm G}^{\mbox{\ttiny \sc (0)}}_{ijk} (\eta)
\eea
Here, without loss of generality, we placed the bulk operator at the center $X=0$, and 
\bea
\label{etad}
\eta = \frac{(x_1-x_2)^2}{(1 + x_1^2) (1+ x_2^2)}
\eea
is the cross ratio.
Assuming the bulk theory is weakly coupled, the leading order contribution to the three point function would be expected to take the form of a tree level Witten diagram. In particular, bulk locality requires that it should be free of any non-local singularities.

The function G${}^{\mbox{\ttiny \sc (0)}}_{ijk} (\eta)$ was computed for example in  \cite{HKLLint, KL15} and found to be given by 
 \bea
 \label{globalb}
 \qquad G_{ijk}^{\mbox{\ttiny $(0)$}}(\etat) 
 \is C_{ijk} \; 
 \etat^{h_k} ~_2F_1 \bigl(h_k + h_{ij},h_k -h_{ij},2h_k;\etat \bigr)
 \eea
 with $h_{ij} = h_i-h_j$. This expression has a branch cut at $\eta>1$.
The value $\eta=1$ corresponds to the configuration where $x_1$ and $x_2$ are antipodal points relative to the location of the bulk operator
\bea
\eta \to 1 \qquad {\rm for} \qquad x_1 \to - \frac{x_2}{x_2^2}.
\eea
For general locations $X$, the branch cut singularity appears where $x_1$ and $x_2$ are at the beginning and end-point of a geodesic through $X$.

It is expected that this non-locality can be removed via an appropriate modification of the leading order operators $\Phi^{(0)}$, that includes the effect of bulk interactions. 
It turns out that holographic CFTs naturally have the required structure that allows one to implement this program. 
Using the CFT bootstrap, it can be shown that 
for every pair of light operators $\calo_i$ and $\calo_j$ in a holographic CFT, there exists an infinite series of double trace operators $\calo_n =[\calo_i \calo_j]_n$ with conformal dimensions $h_n = h_i + h_j + n$ with $n=$ integer, up to corrections of order $1/N$  (For simplicity, we are ignoring spin.) \cite{Heemskerk:2009pn, Fitzpatrick:2012yx, Komargodski:2012ek, Fitzpatrick:2014vua}. The OPE between the two light operators takes the form of  a generalized free field expansion, plus subleading interaction terms 
\bea
\label{opex}
{\cal O}_i(x) \li \calo_j\ra\! & \! \simeq \! & \! \sum_n C_{ijn} \, x^{2n} \, \li\calo_n\ra + \, \frac 1 N \, \sum_k C_{ijk}\, x^{2(h_k-h_i-h_j)}\, \li \calo_k\ra 
\eea
plus descendents. Here we extracted an explicit factor of $1/N$ from the OPE coefficient between three light single trace operators.

The procedure developed in \cite{HKLLint, KL15, Kabat:2012av, Kabat:2012hp, Kabat:2013wga} for obtaining the $1/N$ corrected bulk operator $\Phi$ is to write $\Phi~=~\Phi^{\mbox{\ttiny $(0)$}}$ + an infinite tower of free HKLL operators 
associated to the double-trace operators~$\calo_n$. The coefficients in this expansion are then fixed by the requirement that the branch cut at $\eta=1$ cancels out in the three point function. As an alternative derivation, one can fix the interaction terms by requiring that the correlation functions of bulk operators $\Phi(X)$ continue to satisfy the bulk wave equation in general semi-classical backgrounds  \cite{KL15, Heemskerk:2012mn}. The underlying mechanism is that, after replacing the CFT operators by their  expectation value, the interaction terms automatically produce the required deformation of the linearized bulk wave operator.  We will refer to this approach to restoring bulk locality as the HKLL expansion.

Gravity and gauge interactions play a special role, since they dictate that bulk operators must be defined in
a way that respects diffeomorphism and gauge invariance. In gravity, the required modification of local operators is called gravitational dressing.

 \subsection{Virasoro cross-cap operator} \label{SecVirasoroXcap}

Our aim in this paper is to uncover new clues towards finding an intrinsic CFT definition of bulk operators that optimizes the physical requirements of bulk locality, background independence and diffeomorphism invariance. Necessarily, this bulk operator will need to include a form of gravitational dressing. 
From now on we will specialize to AdS${}_3$/CFT${}_2$. We will comment on the higher dimensional case in the concluding section.

Given that 2D CFTs have enhanced conformal symmetry,  it is natural to investigate what happens if one promotes the definition of bulk operators to the full Virasoro cross-cap states
\cite{HV}. 
There are several pieces of evidence in support of this proposal. First,  Virasoro cross-caps are true geometric defects, that affect the complex structure and orientability of the 
2D space-time. The associated  CFT operator transforms covariantly under the full group of analytic coordinate transformations $(z,\bar{z}) \to (w(z),\bar{w}(\bar{z}))$. As we will see in section \ref{uniformization}, this activates the powerful uniformization theorem as an organizing structure for analyzing CFT correlation functions, in a manner that directly mirrors the geometric properties of the AdS bulk. In  \cite{HV}, this fact was used to show that the matrix element of the cross-cap Ishibashi state between two heavy states matches with the mode function of a bulk field in the corresponding BTZ black hole geometry, see also \cite{daCunha:2016crm}.  In this paper, we would like to explore this proposal further.
We will use the cross-cap states both as guidance for how to think about gravitational dressing, and as an investigative tool for studying bulk dynamics and backreaction from a pure CFT perspective.

Inserting a cross-cap makes the 2D space-time non-orientable.
To describe a CFT on a non-orientable space-time $\Sigma$, it is convenient to view $\Sigma$ as a $\mathbb{Z}_2$ quotient of an orientable space-time $\,\widehat\Sigma$, called the {\it orientation double cover}, or {\it Schottky double} of $\Sigma$. The covering space  $\,\widehat\Sigma$ has two copies of each point on $\Sigma$, one for each orientation, and thus admits a $\mathbb Z_2$ involution that interchanges the two orientations. 
Starting from the CFT on $\,\widehat\Sigma$, we obtain
the CFT on $\Sigma = \widehat{\Sigma}/\Z_2$ by moding out by the orientation reversing $\mathbb{Z}_2$ identification.

Thanks to the uniformization theorem, the Schottky double admits a complex coordinate system such that the orientation reversing involution
takes the form of a M\"obius transformation $ z  \to (az+b)/({cz+d})$ with $ad-bc=1$. We can choose the coordinates $z$ so that the boundary that connects the two sheets of the cover 
takes the shape of a circle
\bea
\label{circle}
(z-x) (\bar{z}-\bar{x}) = y^2.
\eea
Via its center location $(x,\bar{x})$ and radius $y$, the circle specifies a point in AdS${}_3$.
We introduce  the orientation reversing global conformal transformation ({\it c.f.} equation (\ref{globalmap}) )
\bea
\label{xcapzero}
 z \ \to\  \tildez  \! \is\!  x - \frac{y^2}{\bar{z}-\bar{x}}, 
\eea
which interchanges the two sheets of the double cover. The physical space-time $\Sigma$, defined as the quotient under the $\mathbb{Z}_2$ identification $z=\tildez$, is now 
non-orientable: it has  a cross-cap inserted at the center of the circle. 
The simplest example of the above construction produces the real projective plane $\RP^2$, which can be viewed as a cross-cap glued into a sphere, or 
 as the quotient of the sphere by the antipodal map, $\RP^2 = S^2/\Z_2$.
 
A cycle that connect any point $x$ to its $\Z_2$ image represents a non-contractibe cycle of Schottky double of the cross-cap. 
 The proposed definition of the local bulk field, put forward in \cite{HV},  is to identify $\Phi(X)$ with the operator that creates the cross-cap defect (\ref{xcapzero}) and projects onto the Virasoro representation labeled by  $h$ in the corresponding dual channel.  Via the operator state correspondence, the operator $\Phi(X)$ is defined via the reflection condition 
\bea
\label{treflect}
\bigl(\spc T(z) - {T}'(\tildez)\spc \bigr)\, \li\spc \calO(\xX) \ra\! \is \! 0, \\[2.5mm]
{T}'(\tildez)\,d \tildez^2 \, = \, {\bar{T}}(z) d\bar{z}^2 \!\!\!\!\!\!\!
\eea
with $\tildez$ given in equation (\ref{xcapzero}). We can translate the state $|\Phi(X)\rangle$  to the state $|\Phiz\rangle$ placed at the center of AdS${}_3$ by acting with
the global isometry transformation 
\bea
g(X)\, = \! &\!\! & \!\!\!\!\!\! e^{-i t H } e^{\varphi \smpc  l_0} e^{\frac \rho 2\smpc (l_{-1} - l_{1})},\\[2.5mm]
{\rm where:}  \qquad H = L_0 + \bar{L}_0, \qquad
 l_0 = L_0 - \bar{L}_0, \!\!\!\! & &  \ \   \ l_{1} = L_{1} - \bar{L}_{- 1}, \qquad \ l_{-1} = L_{-1} - \bar{L}_1.\qquad\qquad\quad
\eea
When expanded in Laurent coefficients,  equation (\ref{treflect}) implies the infinite set of conditions
\bea
\label{ishib}
\qquad \qquad \qquad \bigl(L_n- (-1)^n\overline{L}_{-n}\bigr) \li \Phiz\ra = 0,\qquad \quad n\in \mathbb{Z}
\eea
which define a cross-cap boundary state. If we ignore bulk self-interactions, it is reasonable to require that the state $|\Phi\rangle$ is spanned by descendents of the primary state associated to the CFT operator ${\cal O}(x)$. Equation (\ref{ishib}) then specifies a unique Ishibashi state $|\Phi\rangle = |h \rangle \! \rangle$ 

The Ishibashi conditions (\ref{ishib}) are necessary to ensure that the cross-cap state are invariant under arbitrary conformal transformations that preserve the location of the circle (\ref{circle}). As we will see in section \ref{SecBackInd} and Appendix \ref{AppWI}, this invariance implies that the correlation functions of $\Phi(X)$ satisfy conformal Ward identities, that in the bulk theory can be interpreted as recursion relations similar to the soft-graviton theorems. The full conformal invariance is also a requisite for the uniformization theorem,
which will play a key role in demonstrating background independence of the bulk operators.

This Virasoro Ishibashi state can be generalized to include gauge fields. AdS${}_3$ bulk theories with gauge interactions correspond to CFT${}_2$ with chirally conserved Kac-Moody currents $J(z)$ and $\bar{J}(\bar{z})$. Ward identities in the CFT give recursion relations between correlation function between bulk operators and boundary currents,
analogous to soft-photon theorems. The absence of non-local branch cuts in these mixed correlation functions  requires extending the set of Virasoro Ishibashi condition with an infinite set of Kac-Moody Ishibashi conditions 
$
\bigl( J_{n} - (-1)^n \bar{J}_{-n}\bigr)\spc \li \Phiz\ra = 0.$
More generally, if the bulk contains massless higher spin fields, the Ishibashi conditions must be imposed for the complete extended chiral algebra of the CFT.
This corresponds to `higher spin dressing' of the bulk operators, and ensures that all gauge constraints and associated Ward identities are satisfied.

\smallskip
 
The proposed identification of cross-caps with bulk fields receives combinatorial support via the counting of complex structure moduli.  The orientable double cover of a sphere $\Sigma_g$ with $g$ cross-caps is a Riemann surface with $g$ handles.
The moduli space of conformal structures of $\Sigma_g$ is the subspace of the moduli space of $\widehat \Sigma_g$ which preserves the $\mathbb Z_2$ involution.
It has real dimension $3g-3$ for $g>1$. Adding a cross-cap adds three real moduli, given by the location and size of the new cross-cap. It also adds a new element to the fundamental group of the surface, along which one can insert a projection onto a given conformal family.
The three moduli are interpreted as the locations of the corresponding bulk operator.
The non-orientable surface with $g=1$ is the Klein bottle. It has one real modulus, which we will interpret as the distance between two bulk operators. In section \ref{sec2ptbulk}, we will use this identification to compute the bulk-to-bulk two point function.

In the next sections, we will provide evidence that the replacement of the global cross-cap state by the Virasoro cross-cap state incorporates the physical effect of gravitational dressing.  Other bulk interactions, such as self-interactions among bulk scalar fields, will require a further modification, which we will outline below. 

\subsection{Holographic cross-cap operator} \label{SecHoloXcap}
 
 \def\Pphi{{\mbox{\scriptsize $\Phi$}}}
 \def\Ppsi{{\mbox{\scriptsize $\Psi$}}}

We  now introduce a more complete CFT definition of local bulk operators, that includes the effect of other bulk interactions besides gravity. Our proposal can be viewed as  a natural CFT implementation of the HKLL construction that removes non-local branch cuts in the bulk-to-boundary three point function $\langle \Phi \calo \calo\rangle$.
An analogous investigation was recently  performed in \cite{NO2}. The main difference between our approach and their set-up is that we take the Virasoro cross-cap states as a starting point, and moreover, we replace the strong conformal bootstrap condition considered in \cite{NO2}  by a weaker bootstrap constraint. We will call this weaker condition the `holographic bootstrap constraint' and the solutions `holographic cross-cap operators'. 

We start from the Ansatz that holographic cross-cap states can be expanded as an infinite weighted sum of Virasoro Ishibashi states 
\bea
\label{hkll}
\li\spc \Phiz \ra \! \is\! \li \spc {h} \spc \ra\!\nspc \ra \; +\,
 \sum_{h_{p}\spc > h } \; \PPhi_p \, \li \spc {p} \spc \ra\!\nspc \ra . 
  \eea
The extrapolate dictionary $\lim_{y\to 0} \, y^{-2h}\, \Phi_h(y,x) = \calo_h(x)$ 
requires that the Ishibashi state $| h\rangle\!\rangle$ appears with unit coefficient $\Phi_h=1$ as the lowest term in this expansion. Higher terms with $h_p> h$ will be automatically suppressed in the $y\to 0$ limit. Moreover,  following HKLL, we will assume that all terms $\PPhi_p$ with $h_p>h$ are of order $1/N$. This is a reasonable assumption, given that the role of the these higher coefficients is to incorporate the effect of bulk interactions.

We wish to identify a natural CFT principle that uniquely fixes the $\PPhi_p$ coefficients.
Consider the matrix elements (c.f. equation (\ref{threept}))
\bea
\label{matrixelts}
  \la \calo_i   \ri \spc \calo_j(x)\spc   \li \smpc {p} \smpc \ra\!\nspc \ra \is \eta^{h_i-h_j} G_{i j p}(\etat) , \qquad \qquad \eta\spc \equiv \spc \frac{1}{1+ x^2},\\[2.5mm]
  \label{threeexp}
  \la \calo_i\ri\, \calo_j(x)\,  \li \Phiz\smpc \ra \is  \eta^{h_i-h_j}  G^{{}^{{}_\Pphi}}_{ij}(\eta),\qquad\qquad
G^{{}^{{}_\Pphi}}_{ij}(\eta) \, = \, \sum_{h_p\geq h} \; \PPhi_p \,\spc  G_{ijp}(\etat) .\qquad,
\eea
which represent three point functions of two local operators and one cross-cap state at $X=0$.  
 Both cross-cap states implement the $\Z_2$ identification $x \leftrightarrow x'= -1/\bar{x}$ for the stress tensor. This involution acts on the cross ratio via $\eta \leftrightarrow 1-\eta$.

As we will discuss in section \ref{SecCorr}, the three point functions $G_{ijk}(\eta)$ can be evaluated via  the method of images, by going to the double cover $\widehat{\Sigma}$ of the non-oriented surface $\Sigma$ with the cross-cap. This method identifies the three point function  $G_{ ij p}(\etat)$ with the chiral Virasoro conformal blocks defined on the double cover\\
\bea
\label{bblock}
\eea
\begin{figure}[h!]
\begin{center}
\vspace{-24mm}
\begin{tikzpicture}[scale=1]
         \draw (-.5,0) -- (.5,0);
          \node  at (0,-.2) {\small ${p}$};
         \draw (-.5,0) -- (-1.1,.6);
          \node [above left] at (-0.9,.55) {\scriptsize $\calo_i(\eta)$};
          \draw (.5,0) -- (1.1,.6);
                    \node [above right] at (1,.5) {\scriptsize ${\calo}'_{\bar j}(1)$};
          \draw (-.5,0) -- (-1.1,-.6);
          \node [below left] at (-0.9,-.5) {\scriptsize ${\calo}_j(0)$};
          \draw (.5,0) -- (1.1,-.6);
           \node [below right] at (1,-.5) {\scriptsize ${\calo}'_{\bar i}(\infty)$};
           \node at (-6.5,0) {\qquad $ G_{ijp}(\etat)\; = \;  \la {\cal O}_{i}(0) {\cal O}_{j}(\eta) \spc {\P}_{p}\spc {\cal O}'_{i}(1) {\cal O}'_{ j}(\infty)\ra \; = \!\!$};
              \end{tikzpicture}
              \vspace{-5mm}
    \end{center}
\end{figure}

\noindent
Here $\P_p$ denotes the projection onto the conformal sector labeled by $p$, and we absorbed the OPE coefficients into the definition of the conformal block. We can think of this chiral four point function as obtained by factorizing the operators ${\cal O}_i$ and ${\cal O}_j$ into their chiral halves,
and placing the right-moving half of each operator at the $\Z_2$ image point of the left-moving half under the orientation reversing $\Z_2$ involution. The $\Z_2$ symmetry fixes the cross ratio $\eta$ to be real.
 
 The bootstrap condition is a restriction on the $\Z_2$ transformation of the three point function. Since the Ishibashi state $| \spc {p} \spc \rangle\!\rangle$ involves a projection on a given conformal sector, correlation functions involving primary fields are not single valued on the quotient space $\Sigma$: if we transport $\calo_i$ from a point $x$ in the outside region $|x|>1$ to its mirror point $x'$ in the interior region $|x|<1$, we must deform the contour along which the projection acts. As a result, the  three point functions $G_{i j p}(\etat)$ exhibit a non-trivial $\Z_2$ monodromy. 
The standard bootstrap condition for CFT cross-cap states includes the requirement that the corresponding three point functions are single valued under the $\Z_2$ involution $
G_{ij}(1-\eta) = \epsilon \spc G_{ij}(\eta)$ up to an overall sign $\epsilon =\pm 1$ \footnote{This condition can be thought of as a generalization of \eqref{treflect} to other non-chiral operators, it can also be thought of as $[\calo_j-\calo'_j]|C\rb=0$, where $\calo'_j$ is the image of $\calo_j$ under the cross-cap identification.}. However, as pointed out in \cite{NO2}, this condition appears to be too restrictive. We will therefore replace it with a weaker requirement that there must exist a mirror cross-cap state $|\widetilde\Phi\rangle = \sum_k \widetilde\PPhi_k \li k\rangle\!\rangle \in {\cal H}_{CFT}$  such that the $\Z_2$ involution  $x \to x'$ interchanges the respective three point functions 
\bea
{G}^{{}^{{}_{\spc \Pphi}}}_{ij}(1-\eta) \is \widetilde{G}^{{}^{{}_{\spc \widetilde\Pphi}}}_{ij}(\eta).
\eea 
In other words, there should exist a set of expansion coefficients $\widetilde\PPhi_k$ such that
\bea
\label{holoboots}
\sum_p\, \Phi_p \spc\spc G_{ijp}(1-\eta) \is \sum_k\, \widetilde\PPhi_k\spc \spc G_{ijk}(\eta),
\eea
where  the sum on the right-hand side must run over the actual CFT spectrum. 
Via the identification (\ref{bblock}), the holographic bootstrap condition (\ref{holoboots}) can be diagrammatically represented in the form of a crossing relation

\smallskip

\bea
\eea
\vspace{-25mm}

\begin{center}
\includegraphics[scale=0.45]{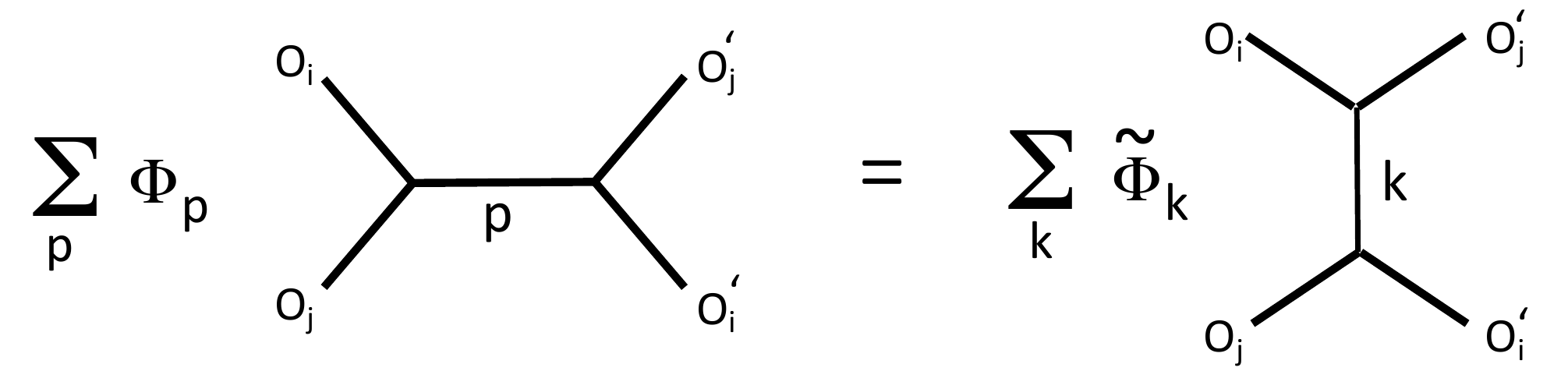}\end{center}
\noindent
This condition can be explicitly analyzed once we know analytic expressions for the Virasoro conformal blocks.

While equation (\ref{holoboots}) is weaker than the standard CFT bootstrap  constraint, it is still very restrictive. Using the identification (\ref{bblock})  and known results about the crossing relations of Virasoro conformal blocks \cite{ponsotteschner}, it may be possible to write a formal duality relation of the form $G_{ijp}(1-\eta) = \sum_k M_{pk} G_{ijk}(\eta)$. However, for holographic CFTs with $c\gg 1$, this duality transformation necessarily involves an integral over a continuum of states, that do not all belong to the CFT spectrum \cite{ponsotteschner}. The rule that the sum  runs over the actual CFT spectrum is a necessary requirement if we want to be able to view the state $|\Phiz\rangle$ as obtained by imposing some geometric boundary condition on the CFT path-integral. Other than that, it is reasonable to assume that the dual expansion coefficients $\widetilde{\PPhi}_k$ describe some random distribution supported on the entire spectrum over the CFT and that all  $\widetilde\Phi_k$  are of order $O(1/N)$.
A further discussion of equation (\ref{holoboots}) from an algebraic (rational) CFT perspective is presented in Appendix A.

\def\ij{{}}

\subsection{Bulk interactions and locality II} \label{SecIntLoc2}

We will now argue that for holographic CFTs, the holographic bootstrap condition (\ref{holoboots}) is sufficient to ensure that the bulk to boundary correlation functions obey micro-causality. 
The HKLL micro-causality condition  stipulates that  the three point function should be free of non-local branch cuts.
This is a special case of the more stringent condition that the three point function remains regular when the operator $\calo_j$ approaches the $\Z_2$ image $\calo'_i$ of
the other operator
\bea
\label{gregular} 
G^{{}^{{}_\Pphi}}_{ij}(\etat)  \is  \textbf{regular at}\ \eta  = 1.\eea
In the dual AdS interpretation, the point $\eta=1$ corresponds to the situation where $\calo_i$ and $\calo_j$ are situated at antipodal points relative to the bulk operator $\Phiz$. 
The condition (\ref{gregular}) is highly restrictive. As we will see, for generic cross-cap states the three point function $G^{{}^{{}_\Phi}}_{ij}(\etat)$ is expected to exhibit a non-local logarithmic branch cut for $\eta>1$. 
Equation (\ref{gregular}) demands that these branch cuts cancel out in the sum~(\ref{hkll}). We will argue how our proposal \eqref{holoboots} satisfies \eqref{gregular}, but we leave for future work to understand if there is a better explanation purely from the CFT point of view.

The regularity of the 3-point function at $\eta=1$ gives useful information about the expansion coefficient $\widetilde\PPhi_k$  in the crossed channel.  As mentioned in section \ref{SecIntLoc1}, the spectrum of a holographic CFTs contains for every pair of light single trace operators $\calo_i$ and $\calo_j$ an  infinite series of double trace operators $\calo_n$ with conformal dimension $h_n = h_i + h_j + n$ with $n$ a non-negative integer. At large $N$, these operators give the dominant contribution in the OPE expansion between $\calo_i$ and $\calo_j$. Using equations (\ref{opex}) and (\ref{holoboots}) we deduce that to leading order in $1/N$, we can expand\footnote{Here we assume that all $\tilde{\PPhi}_p$ coefficients that appear in (\ref{holoboots}) are of the same order in $1/N$. Note that, for the three-point function to be order $1/N$, the coefficients $\tilde{\Phi}_n$  of the double trace operators $\calo_n$ have to be of order $1/N$.  Since all other OPE coefficients are suppressed by an extra factor of $1/N$, the dots thus refer to terms of order~$1/N^2$.} 
\bea
G^{{}^{{}_\Pphi}}_{ij}(\etat) \;
&\!\!\!\!\! \!   \raisebox{-6pt}{$\mbox{$\simeq $} \atop {\eta \to 1}$} \!\!\!\!\! & \;\, \sum_n C_{ijn} \, \widetilde\PPhi_n \,  (1-\eta)^{n} \,
\; + \; ...
\eea
where the sum over $n$ runs over all intermediate double trace operators $\calo_n$.  All leading terms in this sum are regular at $\eta=1$ and in particular have no branch cut at $\eta>1$.   Regularity at $\eta=1$ requires that all subleading contributions of single trace operators and heavy states, indicated by the dots, are suppressed by extra factors of $1/N$ and/or by kinematics. As we have argued above, holographic CFTs are expected to satisfy this requirement.

\begin{figure}[t]
\begin{center}
\includegraphics[scale=0.4]{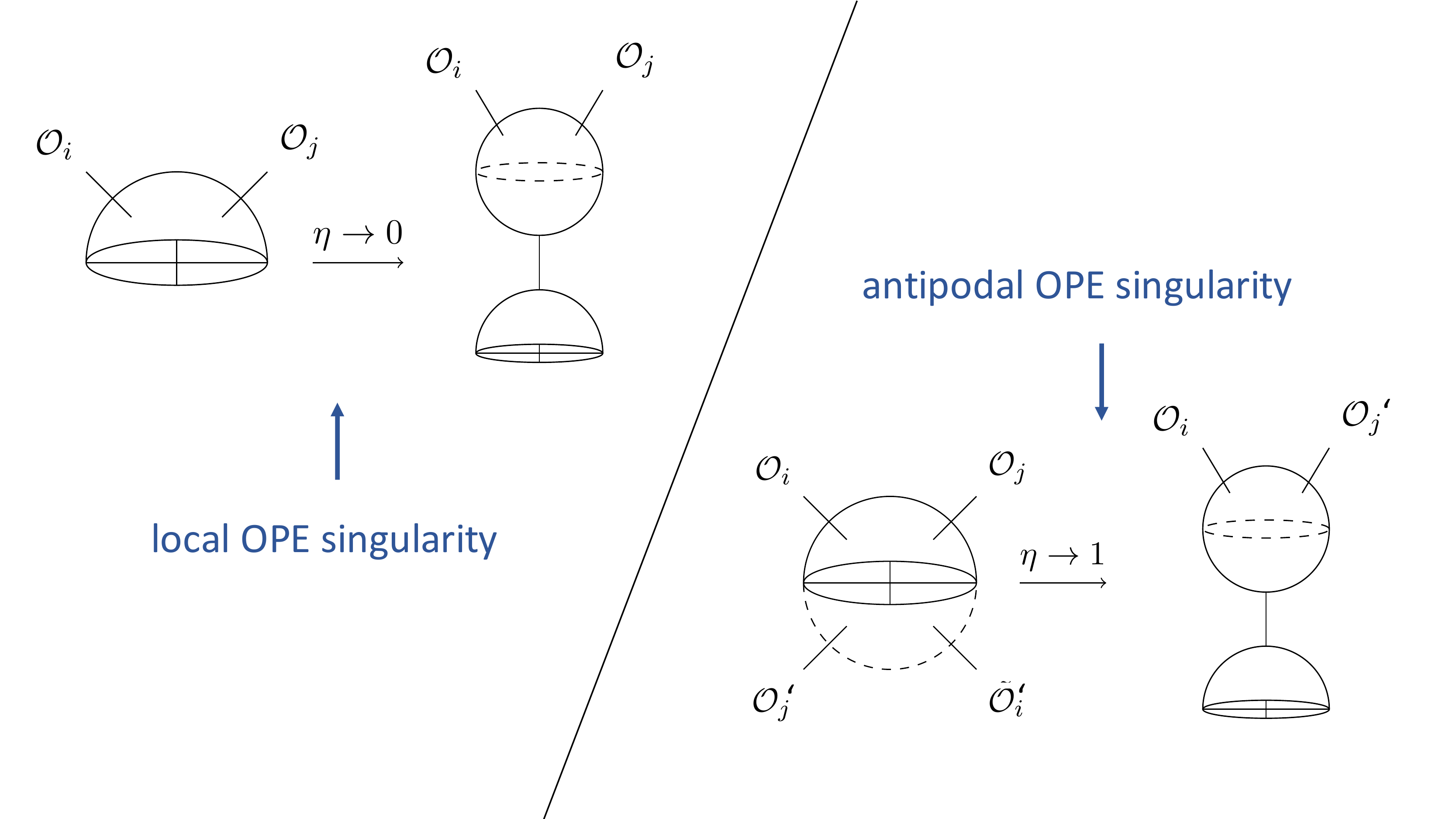}~~~~~~~~~~~~~~~
\raisebox{.5cm}{\includegraphics[scale=0.18]{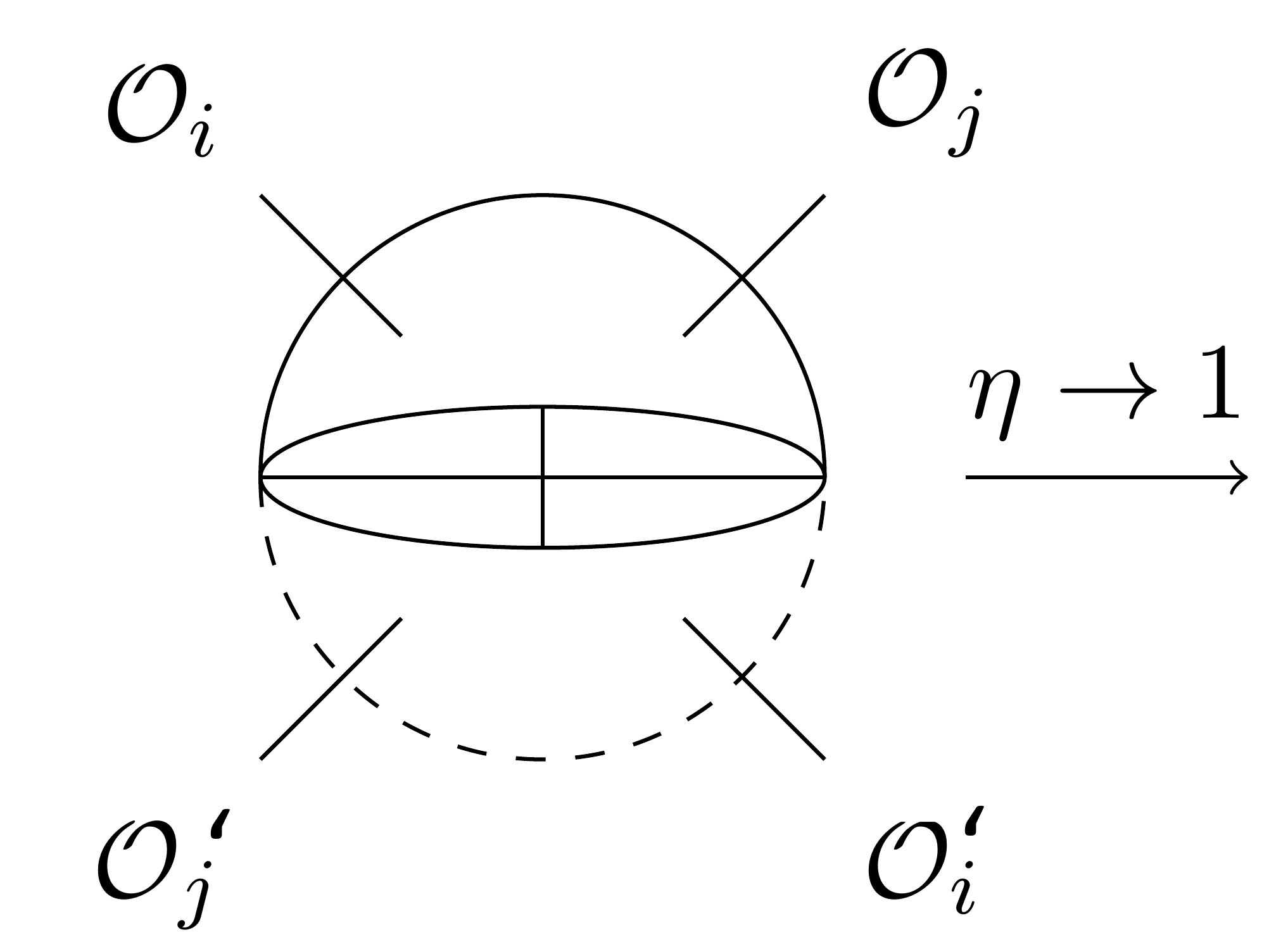}}\includegraphics[scale=0.4]{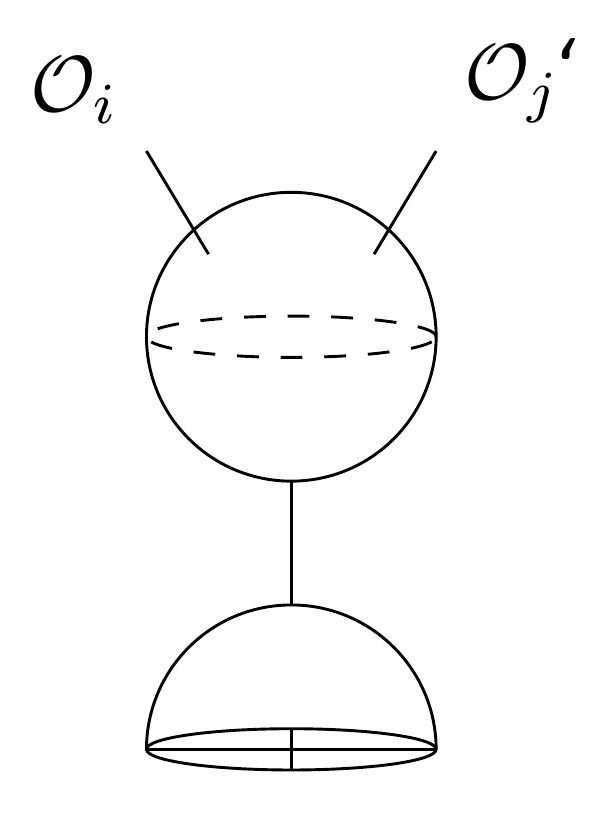}
\caption{\addtolength{\baselineskip}{-.5mm}The limits $\etat \to 0$ and $\etat\to1$ of the bulk-to-boundary three-point function $\la\Phi\calo_i\calo_j\ra$  correspond to the two OPE limits where ${\cal O}_i$ approaches ${\cal O}_j$  or its mirror image \cite{nonocft}. For physical cross caps both limits correspond to the same physical situation. For holographic cross caps, bulk locality requires that the limit $\etat\to1$ is regular.}
\label{eta-limit} 
\end{center}
\vspace{-2mm}
\end{figure}

The regularity condition (\ref{gregular}) was used in \cite{HKLLint} to (uniquely) determine the $1/N$ corrections to $\Phi^{(0)}$. 
We can make direct contact with the analysis of \cite{HKLLint}, by noting that in the large $c$ limit with fixed $h_i$, $h_j$, $h_p$,  the Virasoro blocks reduce to global blocks
\bea
 \label{globalbb}
 \qquad G_{ijp}(\etat)  \raisebox{-6pt}{$\mbox{$= $} \atop {c \to \infty}$}\;  C_{ijp} \; 
 \etat^{h_p} ~_2F_1 \bigl(h_p + h_{ij},h_p-h_{ij},2h_p;\etat \bigr)
 \eea
 with $h_{ij} = h_i-h_j$. This expression matches the bulk-to-boundary three point function (\ref{threept}) of $\Phi^{\mbox{\ttiny $(0)$}}$ found in \cite{HKLLint}. The global block (\ref{globalbb}) has a logarithmic branch cut for $\eta>1$. The CFT origin of this branch cut is that, due to the projection on a fixed conformal family in the intermediate channel, the conformal block has non-trivial monodromy when $\calo_j$ circles around  $\calo'_i$.
 
Let us denote the discontinuity of $G_{ijk}(\eta)$ across the branch cut by $I_{ijk}(\eta)$.  As we have argued above, for holographic CFTs and for bulk operators that satisfy the bootstrap condition (\ref{holoboots}), this discontinuity cancels to leading non-trivial order in $1/N$. So we deduce that the coefficients $\PPhi_p$ in the expansion (\ref{hkll}) must be chosen such that
\bea
\label{hkllbranch}
 \sum_{p} \PPhi_p\, C_{ijp} I_{ijp}(\eta) \! \is\!  0.
 \eea
This equation has been analyzed in \cite{HKLLint} and shown to uniquely fix the $1/N$ corrections, via the 
 Ansatz that the sum over $p$ in (\ref{hkll}) includes (besides the primary state $h$ with normalized coefficient $\Phi_h=1$) an infinite set of double trace operators $\calo_n$ with coefficients $\PPhi_n$ of order $1/N$. After plugging the solution for $\PPhi_n$ into the expansion (\ref{threeexp}) and using (\ref{globalbb}), one discovers that the resulting three point function 
$G^{{}^{{}_\Phi}}_{ij}(\etat)$ reproduces the expression for the tree level Witten diagram of a local interacting bulk theory \cite{KL15}. In combination, these results support our proposal that the holographic cross cap states defined by (\ref{treflect}) and (\ref{holoboots}) exhibit no non-local singularities  (to leading order in $1/N$), and solve the interacting bulk equation of motion.

\medskip

Let us summarize. We have formulated an intrinsic CFT definition of the HKLL bulk operators in terms of cross-cap boundary states. The novelty of our approach is that we factorize the micro-causality conditions into two steps:  (i)~the conformal invariance constraint (\ref{treflect}) and  (ii) the holographic bootstrap equation~(\ref{holoboots}). Both conditions are natural from a CFT perspective, and  do not presume the existence of a bulk dual. We have argued that these conditions are implied by CFT locality, but for holographic CFTs are sufficient to imply that the three point functions $\la \Phi \calo \calo\ra$ are free of non-local branch cuts, at leading order in $1/N$.  However, as we will see in the following, correlation functions involving boundary stress energy tensors and other conserved currents such as $\la \Phi \calo T \ra$ and $\la \Phi \calo T \bar{T} \ra$ 
still exhibit singularities whenever the stress energy insertions are antipodal relative to $\Phi$. Thanks to the conformal invariance constraint (\ref{treflect}) these singularities are given by poles instead of branch cuts. In the next section, we will argue that imposing Virasoro symmetry has a natural bulk interpretation in terms of gravitational dressing.

Before we move on, we would like to point out that a combination of blocks that solves similar constrains in higher dimensional CFT has already been considered long time ago by A. Polyakov \cite{Polyakov:1974gs}, and more recently in \cite{ElShowk:2011ag, Gopakumar:2016wkt}. In \cite{ElShowk:2011ag} they argue that in theories with a spectrum of primary operators and double trace operators the bootstrap equation can be efficiently solved if one first constructs what they called ``dressed" conformal blocks. These blocks are defined as a sum of a single primary operator block plus double-traces in a way that solves the equation \eqref{holoboots} that we called ``holographic bootstrap", although applied to the full global block and not to the chiral part of it. They argue in that paper that this equation has infinite solutions given by sums of double-trace operators that by itself solves \eqref{holoboots}. To pick a special solution they further impose the regularity condition similar to \eqref{gregular}.  A similar logic was initiated in \cite{Polyakov:1974gs} and implemented recently in \cite{Gopakumar:2016wkt} using Mellin transformations techniques. These considerations show that the HKLL states are a natural object that appears when trying to solve the bootstrap in theories with a generalized free field theory-like spectrum. 
\medskip

\section{Gravitational dressing} \label{SecGravDress}

In this section, we  discuss a few basic aspects of gravitational dressing, and formulate some physical expectations about what a dressing operation should look like.
We then study to what extent the Virasoro cross-caps states meet these expectations.

\subsection{Physical expectation}

In gauge theories,  gauge invariance forbids the presence of local charged field operators. It dictates that a charged matter field $\Phi(X)$ must be attached to a non-local operator $W_\Gamma(X)$ that creates the  associated flux line, either in the form of an open Wilson line or as a smeared operator that directly creates the associated Coulomb field.  We will call  $W_\Gamma(X)$  a `dressing operator'.
Dressing operators are not unique. To formulate a notion of locality, it is useful to choose a gauge, that associates a unique dressing operator to a given point. The statement of locality  is that  $\Phi_{\Gamma}(X) = \Phi(X) W_\Gamma(X)$ commutes with operators spacelike separated from the support of $W_\Gamma(X)$.  It is often customary to choose the dressing operator such that in the given gauge choice $W_{\Gamma}=1$ so that $\Phi_{\Gamma}(X)=\Phi(X)$. Note, however, that even if $\Phi_\Gamma(X)$ looks like the usual local QFT operator, the commutator algebra (defined via the Dirac bracket) is non-local, as it includes the non-local effect of the dressing operator.

The gravitational story is quite analogous. In this case, the dressing enforces diffeomorphism invariance and 
creates the gravitational field associated with the excitations created or annihilated by a local operator. A convenient gravitational dressing in AdS space-time is to specify the position of an operator via the affine distance along a geodesic normal the boundary. In Fefferman-Graham coordinates, the associated dressing operator looks trivial. As before, however,  the presence of the dressing operation remains visible in the form of non-local commutation relations  \cite{Donnelly:2015hta}.\footnote{Bulk locality in the presence of gauge interactions has been studied before in \cite{Kabat:2012av,Kabat:2012hp,Kabat:2013wga, Heemskerk:2012np, Guica:2015zpf}. These papers consider bulk fields in the F-G gauge $g_{y \mu}=A_y=0$, which 
makes trivial the dressing defined by shooting a perpendicular geodesic to the boundary of $AdS$. The commutators between the bulk operators and the corresponding conserved currents are found to vanish at space-like separation along the boundary: if a bulk operator $\Phi(y,x)$ is anchored to the point $x$ in the boundary, then $[\Phi(y,x),T_{\mu \nu}(x')]= [\Phi(y,x),j(x')] =0$ for $x\neq x'$. Gravitational dressing in $d=3$ has been explored before in \cite{marolfmintun}. Here we are taking a more operational approach.} 
\def\cK{{\cal K}}

The CFT expression for the geodesically dressed bulk local operator is not known at present.\footnote{While gravitational dressing has been considered before in \cite{Kabat:2012av, Kabat:2013wga, Heemskerk:2012np}, they never wrote an explicit boundary expression with the proper "local" bulk commutation relations.  In the case of electromagnetic interactions, \cite{Guica:2015zpf} analyzed it in detail.}

In general, we expect that the dressing operator receives contributions to all orders in the stress tensor. The following physical argument gives a useful indication for what the result should look like, see \cite{Donnelly:2015hta} for more details. Suppose we choose some gauge, say, by setting some components of the metric equal to zero. Let us denote the coordinate system in this gauge by $\mathcal{X}$. Since diffeomorphism invariance is now fixed, a local field $\Phi^{\mbox{\ttiny (0)}}(\mathcal{X})$ is a physical observable. In some other general coordinate system $X$, we can write the corresponding dressed field as $\Phi(X)=\Phi^{\mbox{\ttiny (0)}}(\mathcal{X}(X))$. Let $v(X)$ denote the vector field that generates the diffeomorphism
$X \to \mathcal{X}(X) + \spc v(X)\spc + \spc ...$ Then the dressed operator looks like
\bea
\label{dressk}
\Phi(X) \is e^{\cK(X)} \spc \Phi^{\mbox{\ttiny (0)}}(X) \, e^{\cK^\dag(X)}, \qquad \qquad
\cK(X) \, =\, i v^{\alpha}(X) \spc P_{\alpha} 
\eea
with $P^{\alpha}$ a bulk translation generator. We choose the coordinate system $X$  such that $\mathcal{X}(X) =X$ in the pure AdS, so that $v^\alpha(X) = 0$ in the vacuum state.  When we deform the background away from pure AdS by turning on some {boundary} stress energy source $T_{\alpha\beta}$, we expect that $v^\alpha(X)$ becomes non-zero, and in the linearized regime, proportional to $T_{\alpha\beta}$. The translation operator $P^{\alpha}$ is also linear in $T_{\alpha\beta}$. We deduce that, in this linearized regime, the dressing operation takes the generic form (\ref{dressk}) with $\cK$ a quadratic expression in the CFT stress tensor
 \bea
  \label{expect}
  \cK \is \frac{1}{N^2} \int \! d^2z \, d^2w\spc f^{\alpha\beta\gamma\delta}(z,w)\spc T_{\alpha\beta}(z)\spc T_{\gamma\delta}(w),
 \eea 
 where  $f^{\alpha\beta\gamma\delta}(z,w)$ denotes a suitable bilocal tensor.
 Here one of the $T_{\alpha\beta}$ factors detects the deformation of the background geometry, and the other one acts as the generator of the diffeomorphism that transforms the free field $\Phi^{\mbox{\ttiny (0)}}(X)$ into the dressed operator $\Phi(X)=\Phi^{\mbox{\ttiny (0)}}(\mathcal{X}(X))$.  The presence of the $1/N^2$ prefactor indicates that the dressing operator incorporates $1/N$ corrections, and thus would appear to become trivial in the large $N$ limit. In the following subsection, we will show that the cross-cap operators are indeed of the expected form (\ref{expect}).

\subsection{Dressing operator} \label{SecDressOp}

The Virasoro cross-cap state decomposes as a sum of a global cross-cap state plus a series of $1/c$ correction terms. We will now analyze these correction terms, and show that they can be interpreted as gravitational dressing: they incorporate the interaction between the bulk operator and the gravitational degrees of freedom in the bulk. As we will see in section \ref{SecBackInd}, they implement a form of background independence: they ensure that $\Phi(X)$ satisfies the bulk wave equation of motion in a general class of background geometries. In this section, we will isolate the gravitational correction terms and combine them into a single `dressing operator'. Indeed, a perhaps somewhat surprising aspect of our proposal is that even the identity operator ${\bf 1}$ has a non-trivial representation as a gravitationally dressed local bulk operator ${\bf 1}(X)$. The following discussion is meant to explain its physical role. We will see that in the end, via a judicial gauge choice, we will be able to set ${\bf 1}(X) = 1$.

It is not hard to convince oneself that there exists a gravitational dressing operation, which we may formally denote by $e^{\cK(X)} =  g(X) \,  e^{\cK}\spc g(X)^{-1}$, such~that
\bea
\li \Phi(X) \ra \! \is \! e^{\cK(X)} \, \li  \Phi^{\mbox{\ttiny (0)}}(X) \ra 
\eea
The dressing operator $e^{\cK}$ consists of a sum of products of Virasoro generators, designed such that it converts a state that satisfies the global cross-cap conditions into a state that solves the Virasoro cross-cap conditions. When acting on the CFT vacuum state, the dressing operator creates the
Virasoro cross-cap state in the identity representation
\bea
\label{idk}
\li {\bf 1}(X)\ra \! \is \! e^{\cK(X)} \li 0 \ra  
\eea
Via the operator state correspondence we can identify ${\bf 1}(X)  =  e^{\cK(X)}$, which we  can think of as the operation of placing the identity operator at the bulk location $X$. This is a non-trivial operation: it inserts a projection operator that restricts the functional integral of the bulk gravity theory to the space of all metrics for which the point $X$ lies at the intersection of an infinite family of geodesics with prescribed endpoints. The functional integral over all bulk metrics with and without this restriction are physically different.

We would like to make this statement more explicit. For simplicity we will work at the linearized level, and specialize to the state $|\mathbf{1}(0)\rangle = e^\cK |\spc 0\spc \rangle$. 
According to the above interpretation, the state $|\mathbf{1}(0)\rangle$ is given by the functional integral over all metrics for which antipodal points, related via $\tildez = - 1/z$, are end-points of geodesics that all go through the center of AdS${}_3$. This restriction on the functional integral is expressed in the CFT via the conditions
$\bigl(L_n - (-1)^n\, \overline{\! L}_{-n}\bigr) \li {\bf 1}(0)\ra = 0.$
Combining these conditions with the Virasoro algebra 
we deduce that to leading order in $1/c$ the operator $\cK$ takes the form
\bea
\label{ksum}
\cK \is \sum_{n=2}^\infty \, (-1)^n \frac{L_{-n} \bar{L}_{-n}}{\frac c {12} (n^3 -n)}_{\strut} \; + \, \ldots
\eea
where the ellipsis indicate higher order terms in $1/c$. Note that the gravitational dressing operation $e^\cK$ looks like a $1/c$ correction: in the infinite $c$ limit,  $e^\cK$ formally approaches unity. As we will see, however, for certain correlators one still needs to work with the exact exponential expression. This is analogous to the electromagnetic case story \cite{Guica:2015zpf}.

 \smallskip
 
\subsection{Bulk interpretation of dressing operator} \label{SecBulkIntDressOp}

We would like to give a bulk interpretation of the gravitational dressing operator $e^\cK$. From the perspective of the stress energy tensor, a Virasoro cross-cap state implements a true $\Z_2$ identification on the CFT geometry. Taking the $Z_2$ quotient produces a non-orientable boundary space-time $\Sigma = \widehat{\Sigma}/\Z_2$. Semi-classical bulk geometries with non-orientable boundaries were recently studied in \cite{geon, MR}. We briefly summarize their results. Following \cite{MR}, we will work in a Euclidean~signature.

Any 3-manifold ${\cal M}$ with non-orientable boundary must be non-orientable itself.  The semi-classical bulk geometry is therefore the $\Z_2$ quotient of an orientable manifold
\,$\widehat{\!\cal M\nspc}$\,.  Unlike its boundary $\Sigma = \widehat{\Sigma}/\Z_2$, the quotient geometry ${\cal M}=\,\widehat{\!\cal M\nspc}\,/\Z_2$  may not be smooth: it has a single $\Z_2$ defect located at its center.\footnote{
The orientable double cover $\,\widehat{\!\cal M\nspc}\spc$ is given by the standard AdS${}_3$ space-time with metric $\widehat{ds}{}^2 = (dy^2 + dz d\bar z)/{y^2}$  in Poincar\'e coordinates,
or $\widehat{ds}{}^2 = \cosh^2\nspc  \rho\spc d\tau^2 + \sinh^2 \nspc \rho\spc d\theta^2 + d\rho^2$, in global coordinates.
The Poincar\'e patch is $y \in \R^+$ and $(z,\bar{z}) \in \C$; the global patch has boundary $S^1 \times \R$ parametrized by $\tau \in \R$  and $\theta \in [0,2\pi]$. The center of global AdS${}_3$ is at $\rho = 0, \tau=0$, which corresponds to $y=1$ and $z=0$. }

The orientation reversing involution $\sigma: \ z \to - 1/\bar{z}$ of the boundary extends to an orientation reversing isometry of the bulk. In Poincar\'e coordinates it reads
\bea
\label{sigmaone}
\sigma: \quad z &  \to & - \frac{z}{|z|^2 + y^2}, \quad y \to \frac{y}{|z|^2 + y^2} , 
\eea
or in global coordinates
\bea
\label{sigmatwo}
\sigma: \quad (t,\rho,\theta)& \to &(-t, \rho, \theta+\pi).
\eea
This transformation has a fixed point at the origin of AdS. The quotient space-time ${\cal M}$ = AdS${}_3/\Z_2$ therefore has a $\Z_2$  singularity at the origin.
We will call this singularity a $\Z_2$ defect.

From the above description it may look like inserting the $\Z_2$ defect amounts to removing half of the bulk space-time. However, 
based on the CFT discussion in section \ref{SecBulkOp} and the correspondence with the HKLL program, we will adopt a different interpretation. We will treat the involution (\ref{sigmaone})-(\ref{sigmatwo}) as an exact symmetry, that is enforced only on all geometric deformations of the bulk space-time. In other words, we require that any deformation of the space-time metric should respect the $\Z_2$ symmetry. From the point of view of the full AdS geometry, this is a global restriction that leads to antipodal correlations in the stress energy tensor. Massless gauge fields and currents are also subject to the same $\Z_2$ symmetry requirement.

The presence of the $\Z_2$ defect has non-local consequences, and thus creating it is a non-local operation. Any sphere surrounding the $\Z_2$ defect is non-orientable, and has non-trivial homotopy. The support $\Gamma$ of the  bulk operator $\Phi_\Gamma(X)$ that creates the defect at $X$ must therefore extend out from the bulk point $X$ all the way to the boundary and insert a cross-cap into every surface that intersects with $\Gamma$. 
This extended support of $\Phi_\Gamma(X)$ also serves to specify the bulk location  of the point $X$, by restricting the bulk geometry such that the (regularized) geodesic distance from $X$ to the boundary is held fixed. This type of non-locality is an inevitable part of any gravitational dressing prescription. 
One of the motivations of our study is to quantify its physical consequences. From here on we will not explicitly indicate the support $\Gamma$ of the bulk operator $\Phi(X)$.

\subsection{Dressing operator as a sum over geometries} \label{SecDressGeo}

The dressing operator ${\mathbf 1}(X)$ is a purely geometrical defect. Note, however, that $\langle T(z)  \mathbf{1}(X)\rangle = 0$, so at the linearized level, the dressing operator leaves the classical metric unperturbed. To see its physical effect, we need to look at the fluctuations of the metric. 
 As we will now show, we can write the state $\li \mathbf{1}(0)\rangle$, defined in equations (\ref{idk}) and (\ref{ksum}),
 as a functional integral over all metrics compatible with the antipodal identification associated with $X$. We will work at leading order at large $c$. In this limit, we can use semi-classical gravity and use a linearized approximation.

In semi-classical gravity, the functional integral over metric is reduced to a sum over classical geometries. 
Consider the general class of vacuum solutions of the 3-d Einstein equations 
\bea
\label{banados}
ds^2 \is \frac{1}{y^2}\bigl(dy^2 \nspc + dz d\bar{z}\bigr) + \tT
\smpc  dz^2 + \overline{\tT}
d{\zb}^2 
+ y^2\spc \tT \, \spc \overline{\!\tT\!} \spc dy^2
\eea
where $\Omega(z)$ and $\bar{\Omega}(\bar{z})$ denote analytic functions of $z$ and $\bar{z}$. Elements within this class of vacuum solutions are all related via reparametrizations  that act non-trivially on the boundary of of AdS${}_{3}$. The continuous family of bulk metrics (\ref{banados}) is associated with the family of CFT vacuum states  $|\Omega \spc 
\rangle= \spc U(\xxi) |0\rangle$ related via diffeomorphisms $U(\xxi) \spc = \spc e^{{\oint \! dz 
\, \xxi(z) \spc {T}(z) + c.c.}\label{tzz}}$, with $\Omega$ and $\xxi$ related via 
\bea
\langle \tT| \smpc T(z) | \tT \rangle \spc = \spc\frac c 6\,  \tT(z) \spc  
= \spc \frac c {12} \partial^3 \xxi . \label{omegaxi}
\eea
These vacuum states $|\Omega\rangle$ are all physically distinct. Their existence is linked with the fact that AdS${}_{3}$ has an infinite asymptotic symmetry group. 
\footnote{Note that the relation (\ref{omegaxi}) between $\Omega$ and $\xxi$ determines the vector field $\xxi$ up to $\mathfrak{sl}(2)$  transformations $\xxi(z) \to \xxi(z) + \epsilon_{-1} + \epsilon_0 z + \epsilon_1 z^2$.
So for given $\Omega(z)$, the vector field $\xxi(z)$ will in general be multi-valued and undergo some $\mathfrak{sl}(2)$  monodromy transformation.}

Now let us consider the state $|\mathbf{1}(0)\rangle = e^\cK |\spc 0\spc \rangle$ created by dressing operator placed at the center of AdS.
From the linearized expression (\ref{ksum}) for $\cK$, we deduce that the state can be written as a gaussian integral over $\Omega$ of deformed vacuum states\footnote{Here $\Omega$ is restricted to have trivial monodromy.}
\bea
\label{sumc}
\li {\bf 1}(0)\ra \is \int [d\Omega]\; e^{- \spc S[\smpc \Omega\smpc] } \, \li \Omega\ra,
\eea
We can uniquely express the exponent $S[\smpc \Omega \smpc]$ in the prefactor as 
\bea
 S[\spc\Omega\spc] \is  \sum_{n\geq 2} \, (-1)^n\spc \frac{c}{12}\spc (n^3 -n)\, \bar{\xxi}_{n} \spc \xxi_{n} \, + \, \ldots\\
\is \frac c 2  \oint \! dw \spc dz
\, \frac{{\xxi}(w) \, \overline{\nspc\xxi}\spc(-1/\bar{z})}{(z-w)^4}  \; + \,\ldots \; 
\eea
with $\Omega$ and $\xxi$ related via (\ref{omegaxi}).
This expression can be recognized as the linearized Einstein action evaluated on the deformed AdS${}_3$ geometry (\ref{banados}), over the past region of the constant time slice, 
and  with the deformation $\Omega$ restricted to be symmetric under the antipodal map
\bea
\qquad S[\spc \Omega\spc ] \is  \frac{1}{8\pi G}\, S_E [\spc \Omega\spc] \\[3mm]
\qquad \Omega(z) \spc dz^2 \is \Omega'(\tildez) \spc d\tildez^2, \qquad \tildez = -1/\bar{z}
\eea
The Newton constant $G$ is related to the central charge $c$ via $\frac 1 {8\pi G} = \frac{2c}{3}$. 
Note that, since the factor $S[\Omega]$ grows linear with $c$, the typical geometries that contribute to the functional integral (\ref{sumc}) are deformed geometries (\ref{banados}) with
$\Omega$ of size $\sim 1/\sqrt{c}$.

\section{Bulk-to-boundary correlators and commutators} \label{SecCorr}

In this section we will compute several correlation functions of a single holographic cross-cap operator $\Phi(X)$ with one or more local CFT operators $\calo(z)$.
These correlation functions are uniquely fixed and computable by the Schottky double construction and conformal symmetry. Consider 
a cross-cap operator located at $X = (y,x,\bar{x})$. To study its correlation functions with local CFT operators, we need to consider the CFT on
the complex plane subject to the $\Z_2$ identification (\ref{xcapzero}). 
We will then analyze the singularity structure of the correlation functions, and deduce information about the commutators between $\Phi(X)$ and local boundary operators.

As it is standard with boundary CFT calculations, see \cite{Difrancesco, nonocft} for example, an arbitrary correlation function with a crosscap Ishibashi state of weight $h_k$ is equivalent to a chiral correlation function in the Schottky double, where we add the appropriate image for each local operator. Now this correlation function can be computed in the plane, with the subtlety that one has to insert a projector ${\P}_{k}$ (into the space corresponding the $h_k$) between the operators and their images.

\subsection{Some bulk-to-boundary correlators}\label{SecBulkBoundary}

In this subsection we will compute some simple bulk to boundary correlation functions. We will work to leading order in $1/N$.

\subsubsection{$\la\Phi \spc {\cal O}\ra$ } \label{SecPhiO}

As a warm-up we start with the bulk-to-boundary 2-point function $\lb \calo(x_1,\bar{x}_1)\Phi(y,x_2,\bar{x}_2)\rb$. 
Local CFT operators $\calo(z,\bar{z})$  factorize into a holomorphic and an anti-holomorphic component. Via the Schottky double, we obtain the bulk-to-boundary correlation functions via the standard image method, by placing a virtual copy of every physical operator at the corresponding mirror point  under the $\Z_2$ identification (\ref{xcapzero}). Specifically, we  map the left moving component at $z$ into a right moving operator placed at $\tildez$. Meanwhile we keep the left-moving component in place.  This procedure also requires a Jacobian factor $\calo(z,\bar{z})\to \frac{y^{2h}}{(\bar{z}-\bar{z}_0)^{2h}} \calo_L(z) \calo_R(\tilde{\bar{z}})$. 
If the operator is already chiral then if it is right moving it will stay at the same location, while if it is left moving it will be mapped to a right moving operator at the mirror point $\tildez$. 

Following this method of images, we  readily obtain 
\bea\label{OPhi}
\lb \calo(x_1,\bar{x}_1)\Phi(y,x_2,\bar{x}_2)\rb \is  \la \spc \calo(x_1,\bar{x}_1) \spc \calo'({x}'_1,{\bar{x}}'_1)\spc \ra
\, = \,  \frac{ y^{2h}}{(y^2+x_{12}\bar{x}_{12})^{2h}}.
\eea
This matches the standard expression for the  bulk-to-boundary propagator. Note that 
if we multiply the result (\ref{OPhi}) by $y^{-2h}$ and take the $y\to 0$ limit, we recover the two point function of a local operator, in accordance to the extrapolate dictionary  $\lim_{y\to0} y^{-2h} \Phi_h =\calo_h$.  The global cross-cap operator would give the same result, as shown by explicit computation in \cite{Takaya1,NO1}.

\def\leftdd{\mbox{\large $\backslash$ }}
\def\rightd{\mbox{\large / }}

  \begin{figure}[t]
\begin{center}
\begin{tikzpicture}[scale=.72]
    \begin{scope}[thick,font=\scriptsize]   
     \path[draw,dashed,blue!50] (1.5,1) circle (3);
    \node [below right] at (.5,.9) {$\mbox{ \normalsize $(y,x,\bar{x})$}$};
     \node [below right] at (-4.7,.7) {$\!\!\!\!\!\! \bullet \; \raisebox{-11pt}{\hspace{-6mm} \normalsize $\calo(z)$}$};
    \node [above right] at (2.8,.9) {$\!\! \bullet \; \raisebox{9pt}{\hspace{-6mm} \normalsize $\calo'(z')$}$};
      \draw [dashed, ->] (1.5,1) -- (-4.5,.5);
     \draw [dashed, ->] (1.5,1) -- (2.7,1.1);
     \path[draw,thick,fill=blue!50] (1.5,1) circle (0.05);
    \end{scope}   
   \end{tikzpicture}
   \vspace{-1mm}
\caption{\addtolength{\baselineskip}{-.5mm} Correlation functions with cross-cap operators are most easily computed by the method of images, by associating to every local operator ${\cal O}(z)$ a $\Z_2$ image operator ${\cal O}'(z')$ placed at the antipodal point $z' = x - \frac{y^2}{\bar{z}-\bar{x}}$.}
\vspace{-4mm}
   \end{center}
   \end{figure}
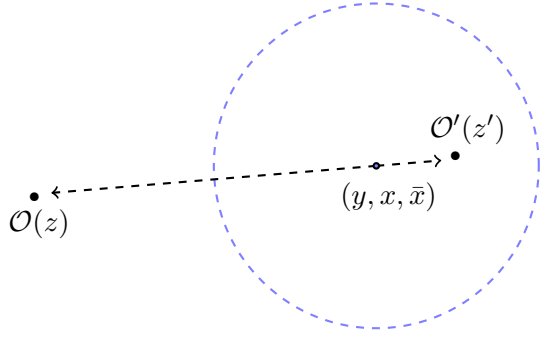

\subsubsection{$\la \Phi \calo T \ra$} \label{SecPhiOT}
Consider the three point function $\lb T \calo \Phi\rb$, where $\calo$ is a scalar operator of the same conformal dimension $h$ as $\Phi$. The Schottky double construction, this is
given by a chiral three-point function on the double cover
\bea\label{TOPhi}
\lb T(z) \calo(x_1,\bar{x}_1)\Phi(y,x_2,\bar{x}_2)\rb \is   \la \spc T(z) \spc \calo(x_1,\bar{x}_1) \spc \calo'({x}'_1, \bar{x}{}'_1)\spc \ra
\nonumber \\[-2mm]\\[-2mm]\nonumber
\is \, \frac{ h~y^{2h}}{(y^2+x_{12}\bar{x}_{12})^{2h-2} (z-x_1)^2 (y^2+ (z-x_2) \bar{x}_{12})^2}, 
\eea
where $x_{12}=x_1-x_2$.  The correlation function (\ref{TOPhi}) exhibits a singularity for $z= x_2 - y^2/\bar{x}_{12}$. The global cross-cap produces the same formula (\ref{TOPhi}). Of course, the reason that the two are identical is that \eqref{TOPhi} follows uniquely from a global conformal Ward identity. 

\subsubsection{$\la \mathbf{1} \,T \spc \bar{T}\ra$} \label{Sec1TT}

As a next simple example, we place a single pure dressing operator $\mathbf{1}(X)$ in the bulk and consider its correlation functions with one or two stress tensor insertions.
We immediately find that these correlation functions are all trivial (i.e. equal to those in the vacuum), except for the $T\bar{T}$  two-point function
\bea
\label{ttone}
\la\spc T(z)\spc \bar{T}(\bar{w})\, {\bf 1}(X)\smpc \ra\!  \is  \frac{c/2}{(y^2+(z-x)(\bar{w}-\bar{x}))^4}
\eea
Note that this two-point function has a pole at $z-x = -y^2/(\bar{w}-\bar{x})$, but that, for $w$ inside the region $|w-x|>y$, the pole in $z$ is located on the second sheet $|z-y| < y$. The presence of the singularity indicates that if we move $T(z)$ from the first to the second sheet, it flips orientation into $\bar{T}(-1/\bar{z})$, which does have a singular OPE with $\bar{T}(\bar{w})$. From the bulk point of view, this antipodal singularity reflects the fact that, in the presence of the dressing operator ${\bf 1}(X)$, metric fluctuations are prescribed to be invariant under the $\Z_2$ involution that keeps the bulk point $X$ fixed. This type of non-locality is a necessary feature of gravitational dressing of local bulk operators.  We will comment further on the necessity and consequences of this non-locality in the next subsection.

\def\hh{h}
\def\hhh{\Delta}
\def\ZZ{\,\spc \mbox{\small \!\!\nspc $\spc Z$}}
\def\mmu{\mu}

\def\Fto{{}_{\mbox{\ttiny 2}}\!\smpc \spc F_{\!\spc \mbox{\ttiny 1}}}

\subsubsection{$\la \Phi \calo T \bar{T}\ra$} \label{SecPhiTT}
The simplest case for which the global and Virasoro results are different is  
\beq
\lb T(z)\bar{T}(\bar{w}) \calo(x_1,\bar{x}_1) \Phi(y,x_2,\bar{x}_2)\rb.
\eeq
In the Virasoro case this is fixed by the image method and the conformal Ward identity. We add the image of the operator $\calo(x_1,\bar{x}_1)$ and the image of the anti-holomorphic component of the stress tensor $T(\tilde{w})$ with $\tilde{w}= - \frac{y^2}{\bar{w}-\bar{x}_2}+x_2$. Including the proper Jacobians we obtain
\bea\label{TTbarV}
&&\lb T(z)\bar{T}(\bar{w})\calo(x_1,\bar{x}_1)\Phi(y,x_2,\bar{x}_2)\rb\, =\, \frac{cy^{2h}}{2(y^2+x_{12}\bar{x}_{12})^{2h}(y^2+z_2\bar{w}_2)^4}\nn
& &~~~~~~~~+\frac{h^2 y^{2h}}{(y^2+ x_{12} \bar{x}_{12})^{2h-4}(y^2+ x_{12} \bar{w}_2)^2(\bar{w}-\bar{x}_1)^2 (z-x_1)^2 (y^2+z_2 \bar{x}_{12})^2} \\[2mm]
&&~~~~~+\frac{2hy^{2h}}{(y^2+x_{12}\bar{x}_{12})^{2h-2}(y^2+x_{12}\bar{w}_2)(\bar{w}-\bar{x}_1)(z-x_1)(y^2+z_2\bar{w}_2)^2(y^2+z_2\bar{x}_{12})}\nonumber
\ea
where $x_{12}=x_1-x_2$, $z_2=z-x_2$, $w_2= w-x_2$, etc. The details of the computation are given in the Appendix \ref{AppTT}. Again we observe that the correlation function (\ref{TTbarV}) is regular everywhere, except for poles when two operators are at the same physical location, or at each others image point on the second sheet.  In particular we see that there are no non-local branch cuts.  

Let us repeat the calculation for the global cross-cap $\Phi^{\mbox{\ttiny \sc (0)}}$. In this case the stress tensor is on equal footing with other quasi-primary operators. Therefore when we insert more than one stress tensor the result is no longer fixed by conformal symmetry. The Schottky double picture is still valid and therefore the three point function is equal to a chiral four point function 
\bea
\lb T\bar{T}\calo \Phi^{\mbox{\ttiny \sc (0)}}\rb \is \lb T\calo ~\P_h^{{}^{{}_G}} ~T \calo\rb,
\eea
 where the first two operator are placed outside the cross-cap radio and the rest are inside and we inserted the projection ${\P}_h^{{}^{{}_G}}$ onto the global module of the primary state associated to the cross-cap. From this one can immediately conclude that this is given by a global conformal block\footnote{In Appendix \ref{AppTT}, we derive this result by direct computation.}
 \bea\label{TTbarG}
\lb h | T(z)\bar{T}(\bar{w})  | \Phi^{\mbox{\ttiny \sc (0)}}\rb\is \frac{h^2}{(-z\bar{w} )^2} ~\Fto\left(2,2,2h;-\frac{1}{\bar{w} z} \right).
\eea
This hypergeometric function diverges at $z\to \tilde{w}=-1/\bar{w}$, which corresponds to the antipodal point on the second sheet, but has a branch cut for $\frac{\tilde{w}}{z} = (1,+\infty)$. This equation can be generalized for arbitrary positions since there is only one cross-ratio that can be defined $\etat = \frac{(z-x_1)(\bar{w}-\bar{x}_1)}{(1+x_1\bar{x}_1)(1+z\bar{w})}$, following \cite{NO2}, since this is equivalent to inserting a non-chiral operator $\calo(z,\bar{w})=T(z)\bar{T}(\bar{w})$. 

Equation \eqref{TTbarG} can be compared with the Virasoro cross-cap result given by \eqref{TTbarV} in the limit in which $x_1,\bar{x}_1\to\infty$ and $x_2,\bar{x}_2\to 0$, which gives 
\beq
\label{ttbv}
\lb h | T(z)\bar{T}(\bar{w})  | \Phi\rb= \frac{h^2}{(-z\bar{w})^2} + \frac{2h}{(-z\bar{w})(1+z\bar{w})^2}+\frac{c/2}{(1+z\bar{w})^4}.
\eeq
Comparing (\ref{TTbarG}) with (\ref{ttbv}), we notice several differences. The global cross-cap correlator has branch cuts, whereas the Virasoro cross-cap correlator has only poles.
On the other hand, equation (\ref{ttbv}) has a term proportional to the central charge $c$, while (\ref{TTbarG}) in independent of $c$. The extra term is identical to the vacuum two-point function $\lb 0 | T(z)\bar{T}(\bar{w}) | 0 \rb$.  Both differences can be interpreted as a result of the gravitational dressing. 

At this stage, it is unclear to us how unique the modification of the global state that gets rid of branch cuts is. For example, it would be nice to understand if the family of 1/c corrections that don't have a branch cut is in one-to-one correspondence with different Virasoro cross-cap states.

For completeness, it is interesting to compare \eqref{TTbarG} and \eqref{ttbv}. Using results from \cite{Osborn:2012vt} we can write the explicit combination 
\beq
\lb h | T(z)\bar{T}(\bar{w})  | \Phi_{(h,h)} \rb_V = \sum_{n=0}^\infty C_n   \lb h | T(z)\bar{T}(\bar{w})  | \Phi^{(0)}_{(h+n,h+n)} \rb,
\eeq
where the coefficients are given by
\bea
C_n &=& \frac{1}{h^2}\left( \frac{c(n-1)_3}{12} (2h)_{n+1} + 2h(n(n+2h-1)+1)(2h-2)_n \right)  \frac{(2h-2)_n}{(2h-2)_{2n+1}} \nn
&&+ \frac{1}{h} (-1)^n( (n+1)(2h-2+n)-2)(n+1)! \frac{(2h-2)_n}{(2h-2)_{2n+1}}.
\ea
This gives another way to write down the Virasoro dressing of the global cross-cap states. The cross-cap states $| \Phi^{(0)}_{(h+n,h+n)} \rb$ are interpreted as Virasoro descendants of $| \Phi^{(0)}_{(h,h)} \rb$. Of course, the sum over double-traces from the HKLL prescription, that are not constructed from Virasoro descendants of the primary, are not going to contribute in this correlator. 

\subsubsection{$\la \Phi {\cal O} {\cal O}\ra$} \label{SecPhiOO}

Next let us consider the three-point function between two local CFT operators and one cross-cap operator. These are proportional to OPE coefficients and should be interpreted as bulk matter interactions. As discussed in section \ref{SecHoloXcap}, in the interacting bulk operators are defined as the holographic cross-cap states $\li \Phiz\ra = 
|h\rangle \! \rangle + \sum_n \Phi_n \li n \ra\!\nspc\ra$, required to solve the equation (\ref{holoboots}). Using the prescription outlined above, we find that the three point function $ \langle \calo_i| \calo_j(x)| \Phi\rangle = \eta^{h_i-h_j} G_{ij}^\Phi(\eta)$  can be decomposed  as $
G^{{}^{{}_\Pphi}}_{ \smpc i\smpc j}(\etat)= G_{ijh}(\eta) + \sum_{n} \PPhi_{ n}\; G_{ i j n}(\etat) ,$ with $G_{ i j p}(\etat)$ given by the conformal blocks defined on the double cover
\bea 
G_{ ij p}(\etat)\hspace{-4.5mm} & & = \;  \la {\cal O}_{i}(0) {\cal O}_{j}(\eta) \spc {\P}_{p}\spc {\cal O}'_{i}(1) {\cal O}'_{j}(\infty)\ra \nonumber \\[3mm]
& & \hspace{-2.5mm}  \raisebox{-6pt}{$\mbox{$= $} \atop {c \to \infty}$}\;  C_{ijp} \; 
 \etat^{h_p} ~_2F_1 \bigl(h_p + h_{ij},h_p-h_{ij},2h_p;\etat \bigr)
 \eea

\noindent
The coefficients $\PPhi_n$ are determined  by the holographic bootstrap condition (\ref{holoboots}), which in particular implies that the function $G^{{}^{{}_\Phi}}_{ij}(\eta)$ has no non-local branch cut for $\eta>1$. As discussed in section \ref{SecIntLoc2},  this requirement is identical to the HKLL prescription for restoring bulk locality \cite{HKLLint}. Via our construction, we give a direct CFT interpretation of the HKLL conditions as the rule that the operator $\calo_j(\eta)$ should have a regular OPE with the antipodal image of $\calo_i$.

\subsection{Commutators} \label{SecComm}

What can we learn from the above explicit formulas for the correlation functions about the commutators between bulk and boundary operators?
So far we have worked in Euclidean signature, where all operators are space-like separated. To extract commutation relations from Euclidean correlation functions we can use the reconstruction theorem of Osterwalder-Schader \cite{Osterwalder-Schader}: we can get arbitrary time ordered correlators by displacing each operator slightly in Euclidean time $O_j(t-i \epsilon_j)$ and then taking the $\lbrace \epsilon_j \rbrace \rightarrow 0$ in a particular order. We can then derive the equal time commutation relations
by extracting the discontinuity at $t_E=0$ 
\bea
 \bigl[\Phi(y,\theta,0),\calo(\theta',0)\bigr]\is \lim_{\epsilon \rightarrow 0} \left ( \, \Phi(y,\theta,\epsilon) \spc \calo(\theta',0) \, -\,   \Phi(y,\theta,- \epsilon)\spc \calo(\theta',0)\,  \right )
\eea 
A nontrivial commutator results only if the field $\cal{O}$ encounters non analyticities, such as poles or branch-cuts, as we vary $x$. 
 Poles are typically associated with commutators between light-like separated operators, while branch cuts
in equal time correlation functions indicate the presence of non-zero commutators at space-like separation. 

Does our proposed definition of bulk operators lead to unacceptable non-local commutators? 

We first look at the c-number part of the commutator $[\Phi(X),\calo(z)]$ between a bulk and a boundary operator. The  2-point function $\la \Phi(X) \calo(z)\ra$ given in (\ref{OPhi}) is, perhaps surprisingly, exactly equal to the free field bulk-to-boundary propagator. So from this we conclude that  $\la [\Phi(X),\calo(z)]\ra = 0$  for $X \neq z.$
So the commutator has no non-local c-number contribution.

Next we look at the commutator $\lb T(z) \mathbf{1}(0) \bar{T}(\bar{w})-\bar{T}(\bar{w}) \mathbf{1}(0) T(z)\rb$ in the presence of a bulk operator $\mathbf{1}(0)$. The $\langle T(z) \mathbf{1}(0) \bar{T}(\bar{w}) \rangle$ three point function given in (\ref{ttone}) has a pole at $w= \tildez =-1/\bar{z}$. This pole gives rise to a non-local commutator when $T(w)$ and $\bar{T}(z)$ are located at antipodal points. For two points to be antipodal and slightly away from $t=0$, the time at which the bulk operator is inserted, they have to the future/past of the crosscap. A similar conclusion holds for the correlation functions $\langle \Phi\calo T \bar{T}\rangle$.
This non-locality appears to be an inevitable consequence of the gravitational dressing operation. Generators of global AdS isometries are given by integrals of the stress energy tensor over the boundary, but act non-trivially on the location of the bulk point. Both facts can be simultaneously true only if there are non-local correlations between the bulk operators and the boundary stress tensor. 

Some partial insight can be gained by studying the Ward identities that express the effect of adding stress energy tensors to correlation functions. From the bulk perspective, these identities are analogous to soft-graviton theorems, that describe the effect of adding a boundary graviton to the correlation function. For a short description of these Ward identities we refer to Appendix \ref{AppWI}. One basic consequence of this bulk interpretation of the Ward identity is that correlation functions of stress tensors cannot have branch-cut singularities, only poles. As we have seen in the previous section, the transition from global to Virasoro cross-cap states indeed eliminates the branch-cut at $\etat>1$, but leaves a pole.

Finally we consider the commutator $[\calo_i(x_1), \calo_j({x_2})]$ in the presence of a holographic cross-cap operator $\Phiz$. The three-point function for Virasoro Ishibashi cross-cap states has a branch cut along the line $\eta>1$. Rotated to Lorentzian signature, $\etat >1$ corresponds to the region $(x_1 - {x'}_2)^2<0$, which includes space-like separated points. The presence of the branch cut would lead to an unacceptable physical commutator between two space-like separated operators. As argued in section \ref{SecIntLoc2}, the holographic bootstrap constraint is sufficient to remove this branch cut and any non-trivial monodromy around $\eta=1$, at least to leading order in $1/N$. This eliminates the non-local commutator. The $\langle \calo_i \calo_j \Phi\rangle$ correlation function continues to be singular at $\eta = 0$ and exhibit a branch cut at $\eta<0$. This is the physical OPE singularity and discontinuity when $\calo_i$ and $\calo_j$ cross each others light-cone.

The gravitational dressing that the cross-cap implements leads to a $\mathbb{Z}_2$ symmetric distribution of energy everywhere on the boundary. It would be interesting to study how, by using conformal transformations similar to those of \cite{marolfmintun}, one could comb the effect of the dressing to a smaller subregion. As we will see in the next section, one indeed has the freedom to change the shape of the cross-caps. This freedom plays an important role in establishing background independence of the bulk operators.

\medskip
\section{Background independence} \label{SecBackInd}

We have seen that the  2-point function $\langle \Phi(X) \calo(x)\rangle$ reproduces the bulk-to-boundary propagator in AdS.
We will now present a general argument that demonstrates that $\Phi(X)$ also satisfies the wave equation in a deformed background. Hence the cross-cap operators provide a concrete realization of an adjustment mechanism, whereby the gravitational dressing terms, when acting on a state that represents a deformed background, automatically produce the necessary modification of the wave equation.

Consider the general class of classical bulk metric  (\ref{banados}).
In the CFT dual, this class of backgrounds correspond to states $|\tT\rangle$ created by acting with a diffeomorphism on the vacuum\footnote{This class of metrics admits a continuous symmetry group generated by the infinitesimal diffeomorphisms
\bea
\label{ctrafo}
z \! &  \to & \! \!  \tilde z \spc =\spc   z + \xxi + \frac 1 4 y^2 \spc { \bar\partial^2\bar{\xxi} }, \qquad \ \ 
 y \, \to \,  \tilde y  \spc =\spc   y \spc \bigr(1+\frac 1 2 (\partial\xxi + \bar\partial\bar{\xxi})\bigr)\
 \eea
under which $\tT$ transforms via $
\tT \to \tilde\tT \spc = \spc \tT    + \xi \partial\tT + 2 \partial\xi \tT - \frac 1 2\partial^3 \xi.$
These infinitesimal diffeomorphisms exponentiate to finite complex conformal transformations on the boundary. The finite transformation law of $\tT$ reads
\bea
\tT(w) dw^2 \is \bigl(\spc \tT(z) - \frac 1 2  \{ w,z\}\bigr) dz^2.
\eea}
$
\li \spc \Omega \spc \ra \, =\, \spc U^\dag(Z) \li 0\ra
$, where $U(Z)$ 
denotes the unitary operator that implements the finite conformal transformation $z \to Z(z)$, with $Z(z)$ chosen such that
\bea
\label{texp}
 \langle \tT| \smpc T(z) | \tT \rangle \spc = \spc\frac c 6\,  \tT(z) \spc  
= \spc \frac c {12}\{ Z,z\}
\eea
Similarly, we also introduce right-moving coordinates $\bar{Z}(\bar{z})$.
We will call $(Z,\bar{Z})$ the uniformizing coordinate system. The 2D uniformizing coordinates can be extended into the bulk to a 3D uniformizing coordinate system $\tilde{\xX}=(\yY,\zZ, \bar\zZ)$, in which the metric (\ref{banados})
takes the standard form
\bea
\label{Zmetric}
ds^2 \is \frac1 {\yY^2} \bigl(d\yY^2 \nspc + d\zZ d\bar{\zZ}\bigr).
\eea

We would like to demonstrate that, when acting on the deformed state $|\tT\rangle$, the bulk field $\Phi(\xX)$ satisfies the wave equation in the corresponding background 
\bea
\label{boxomega}
(\square_\Omega + m^2) \Phi(X) \is 0.
\eea
As we will see,  this statement of background independence is an immediate consequence of the uniformization theorem. Essentially, the CFT proof of (\ref{boxomega}) directly copies the bulk procedure that locally transforms the background metric (\ref{banados}) to the standard AdS${}_3$ form by going to the uniformizing coordinates $\tilde{\xX}=(\yY,\zZ, \bar\zZ)$.  

A special case of this adjustment mechanism was already uncovered in \cite{HV} for the case of the BTZ background. 
For a general metric in the class (\ref{banados}), the uniformizing coordinate
is multivalued: in going around the $S^1$ circle, $Z$ may undergo a monodromy $Z \to \frac{a Z + b}{cZ+d}$ with $ab-cd =1$
and similar for~$\bar{Z}$. For BTZ black hole states, the conjugacy class of the M\"obius transformations specifies the total mass and spin of the black hole \cite{citeBTZ}. 
Specializing to the non-rotating case, $\langle T(z)\rangle={ \Delta}/{z^2}$ and $\langle T(\bar{z})\rangle={ \Delta}/{\bar{z}^2}$, the uniformizing coordinate transformation reads
\bea
\label{uniformz}
\ZZ(z)= z^{ir_{\! +}}, \quad \bar{\ZZ}(\bar{z}) = \bar{z}^{ir_{\! +}}  \qquad {\rm with} \qquad \ \ \ r_{\!+}^2 = \textstyle {\frac{24 \Delta}{c}-1}.
\eea 

The paper \cite{HV} considered the matrix element of the cross-cap operator $\Phi_h(X)$ between two highly excited states with conformal weight $\Delta \gg c/12$. 
This amplitude can be computed via the same method as used in the previous section, by going to the Schottky double.
It is given by the four point conformal block of four heavy external states of dimension close to $\Delta$  projected onto a intermediate channel labeled by the light conformal dimension $h$.
 The $\Z_2$ reflection symmetry of the Schottky double restricts the cross ratio $\eta$  to be a real number, which is identified with the radial location of the bulk point $X$ at which $\Phi$ acts. Setting $X = (\eta,0,0)$, it was found that this matrix element at large $c$ is given by
\bea
\label{btzmode}
 \la h_3,\nspc h_4
\li \Phi_{\hh}(\eta,0,0) \li \spc h_1, \nspc h_2\ra \, =\, \textstyle \eta^h\, \Fto\bigl(h\nspc +\nspc \frac{i}{2r_{\! + }\! }\spc h_{13}, h \nspc +\nspc  \frac{i}{2r_{\! + }\!\!}\, h_{24}, 2\hh\spc ; \eta\bigr) \ \ \  \\[3mm]
\textstyle h_1\nspc \nmpc =\nspc  h_2\nspc\nmpc  = \nspc \frac 1 2  \Delta, \ \ \ h_3\nspc\nmpc = \nspc \frac 12 (\Delta\nspc\nmpc +\nspc \omega\nspc\nspc +\nspc\nspc \ell), \ \ \ h_4
\nspc\nmpc = \nspc \frac 1 2(\Delta \nspc\nmpc +\nspc  \omega\nspc\nspc - \nspc\nspc \ell).\qquad \ 
\nonumber 
\eea
This expression exactly matches with the mode function of a free field of mass $m_h^2 = 2h(2h-d)$ propagating in the BTZ black hole background of mass $M = \Delta - \frac c {24}$. 
Remarkably, the amplitude of the Virasoro cross-cap operator $\Phi(X)$ automatically adjusts itself, so that it solves the wave equation in the appropriate BTZ geometry.
How does this adjustment mechanism work?

\subsection{Gauge freedom}\label{SecGaugeFreedom}

Up to now we have assumed that the $\mathbb{Z}_2$ identification that defines the cross-cap operators acts via an antipodal map along a circle (\ref{circle}) in the local coordinate system. The Ishibashi conditions (\ref{treflect}) imply that the cross-cap states are invariant under diffeomorphisms that leave the circle (\ref{circle}) invariant and that commute with  the antipodal identification $\tildez=z$. The group generated by these diffeomorphisms is isomorphic to Diff$(S^1)$, the diffeomorphism group of the circle.

We can also act with diffeomorphism  that do not respect the invariance conditions. Or in more plain language, we are free to change coordinates.
This would change the equation of the circle. It is then clear that, from the point of view of some fixed coordinate system, there is a continuous infinite dimensional family of cross-cap boundary states related via diffeomorphisms. 

\begin{figure}[h]
\begin{center}
\includegraphics[scale=0.56]{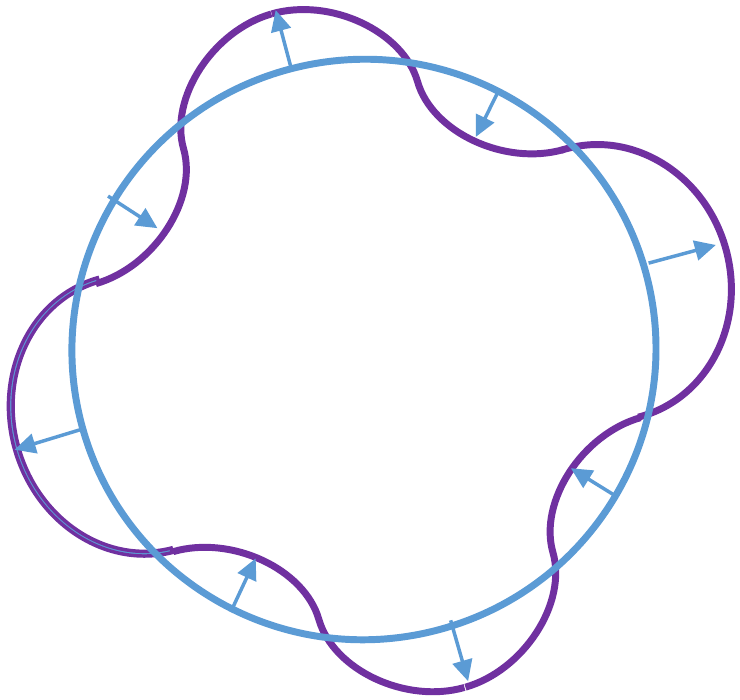}
\caption{\addtolength{\baselineskip}{-.5mm}Deformation of the circle along which the cross-cap identification is made. }
\label{dcircle} 
\end{center}
\vspace{-3mm}
\end{figure}

In general, we can write an orientation preserving diffeomorphism of the form
\bea
\label{xcapg}
 \tildez  = \bar g(\bar z), \qquad \ \ 
\bar g(\bar z) \ = \,  x - \frac{y^2}{\bar{z}-\bar{x}_0}  \, - \,\bar{v}(\bar{z})
\eea
with $\bar v(\bar z)$ some arbitrary anti-analytic function near $z=x$. The original circular boundary between the first and second sheet
is now deformed, as indicated in figure (\ref{dcircle}). The group of deformations of the circle is also isomorphic to Diff$(S^1)$.
Indeed,  in terms of Virasoro generators, the situation is: \\[1mm]
${}$~~(i) the Diff$(S^1)$ generators
$l_n = L_n - (-1)^n \bar{L}_{-n}$\, {(with $n\in \mathbb{Z}$)} preserve the shape of the circle\\[1mm] 
${}$~~(ii) the $\widetilde{\mbox{Diff}}(S^1)$ generators $\tilde{l}_n = L_n + (-1)^n \bar{L}_{-n}$  (with $n\in \mathbb{Z}$)
deform the shape of the~circle.\\[1mm]
Note that the generators from the two different Diff$(S^1)$ groups commute with each other.

We learn that the gravitationally dressed bulk operators $\Phi(X)$ are in fact not defined as functions on AdS${}_3$ but as functionals defined on the group $\widetilde{\mbox{Diff}}(S^1)$. From now on, we will therefore denote them as $\Phi(g)$. 

How can this be compatible with our earlier claim that gluing a cross-cap into a Riemann surface only adds three real  moduli parameters?  The answer is well-known. The  $\widetilde{\mbox{Diff}}(S^1)$ element $g$ can be identified with the transition function, or gluing map, that connects the local coordinate neighborhood of the cross-cap with the complex coordinate system on the remaining part of the Riemann surface.
 The number moduli of the cross-cap is equal to the dimension of a cohomology group: the moduli are the holomorphic deformations of the gluing map that can not be extended
to  holomorphic coordinate transformations on 
 the cross-cap or on the rest of the Riemann surface.   The Riemann-Roch theorem implies that the dimension of this cohomology group is equal to 3 times the Euler character of the cross-cap = 3 $\times$ 1 = $3$.  Below we will summarize this classical argument from the point of view of a CFT correlation function involving the bulk operator $\Phi(g)$.

\subsection{Gauge invariance}\label{SecGaugeInv}

Consider the bulk boundary two point function evaluated between the deformed vacuum state
\bea
\label{twopt}
\la \spc \Omega  \spc \ri \spc \Phi(g)\, {\cal O}(x) \spc  \li \spc \Omega \spc \ra 
\eea
Within this matrix element, the stress energy tensor receives a vacuum expectation value given by (\ref{texp}). We can distinguish two types of gauge transformations\\[1mm]
${}$~(a) the group of conformal transformations that leave the {\it in} and {\it out}
states $\li \Omega\ra$ invariant \\[1mm]
${}$~(b) the group of conformal transformations that leave the  bulk operator  $\Phi[\spc g \spc ]$ invariant.
Both gauge symmetries are active ingredients in the derivation of the uniformization theorem.

A given state $\li \Omega\ra$ satisfies an infinite set of Virasoro conditions of the form $L(\smpc \xi_n) \li \Omega \rangle =  0$, $\bar{L}(\smpc \bar \xi_n) |\Omega \rangle \, =\, 0$, 
where $L(\smpc \xi\smpc ) = \oint \! dz\, \xi(z)\spc  T(z)$ and $\xi_n = z^{n+1} + ...$ 
are a suitable set of vector fields.
Let $V(\xi)$ denote a general element of the conformal group obtained by exponentiating the $L(\smpc \xi_n)$ and $\bar{L}(\smpc \bar \xi_n)$ generators. The Virasoro conditions  imply that the state $\li \Omega \ra$ is inert under $V(\xi)$:
\bea
\label{stateinv}
V(\xi) \li \Omega \ra \! \is \!  \li \Omega \ra. 
\eea
A similar story applies to the bulk operators $\Phi[\smpc g \smpc]$. Each $\Phi[\smpc g\smpc]$ satisfies an infinite set of Ishibashi conditions $
\bigl[L(\smpc \eta_n) , \Phi( g) \bigr] = 0,$
where $L(\smpc \eta) \, = \, \oint \! dz \, \eta(z) \bigl(T(z)  - {T'}(z')\bigr)$  and $\eta_n$  
are a suitable set of vector fields.
Let $V(\eta)$ denote a general element of the conformal group obtained by exponentiating the $L(\smpc \eta_n)$ generators. The Ishibashi conditions then tell us that
the bulk operator $\Phi( g)$ are inert under the action of $V(\smpc \eta\smpc )$
\bea
\label{opinv}
V^\dag(\eta)\spc \Phi( g )\spc V(\eta)\! \is \! \Phi(\spc g \spc) 
\eea

\subsection{Uniformization}\label{uniformization}

We now outline the CFT derivation of the uniformization theorem.  Consider the bulk-to-boundary 2-point function (\ref{twopt}). Both the states $| \Omega\rangle$ 
and the bulk operator $\Phi( g)$ depend on an infinite set of parameters.  The uniformization theorem states that, thanks to the two gauge invariances described above,
we can always find a coordinate system such that the two point function depends only a finite set of parameters, and exactly reduces to the two point function in pure AdS${}_3$.

The gauge invariance (\ref{stateinv}) of the {\it in} and {\it out} states can be used to deform 
the bulk field via $\Phi(g) \to  \Phi( \tilde{g}) \, = \,   V^\dag(\xi)\, \Phi( g ) \,V(\xi)$.
We use this freedom to bring the cross-cap operator in a standard circular form (\ref{xcapzero})-(\ref{ishib}) 
\bea
\quad V(\xi): \quad \Phi(g)  & \to & \Phi(X)
\eea
This standardized bulk operator $\Phi(X)$ still satisfies an invariance property $
V^\dag(\eta)\spc \Phi(X)\spc V(\eta) = \Phi(X)$,
where now $V(\eta)$ denote a general element of the conformal group generated by the Ishibashi generators that annihilate the circular cross-cap state. We can therefore use this invariance to deform the  {\it in} and {\it out} states, and rotate them into the standard $SL(2,\R)$ invariant vacuum state
\bea
\quad V(\eta): \quad \li \Omega \ra &  \to &  \li \spc 0 \spc \ra
\eea

  \begin{figure}[t]
\begin{center}
\begin{tikzpicture}[scale=1.25]
\draw (-1.5,-1) -- (1.5,-1);
\draw (-1.5,1) -- (1.5,1);
 \draw (-1.5,1) arc (90:270:1);
 \draw (1.5,-1) arc (-90:90:1);
  \draw (-1.5,1) arc (180-19.5:180+19.5:3);
   \draw[dashed] (-1.5,1) arc (19.5:-19.5:3);
   \draw (1.5,1) arc (180-19.5:180+19.5:3);
   \draw[dashed] (1.5,1) arc (19.5:-19.5:3);
 \node at (2.05,0) {$|\Omega\rb$};
 \node at (-2.05,0) {$\lb \Omega |$};
 \node at (0,-1.5) {$\calo(x_1)$};
  \node at (0,1.5) {$\Phi(g)$};
  \draw[fill] (0,-0.8) circle (0.05);
  \draw (0,0.8) ellipse (0.2 and 0.12);
  \draw (-0.1,0.8-0.1) -- (0.1,0.8+0.1);
  \draw (-0.1,0.8+0.1) -- (0.1,0.8-0.1);
  \node at (3.4,0) {$=$};
  \draw (6,0) circle (1.2);
   \draw[fill] (6,-1) circle (0.05);
   \node at (6,-1.6) {$\calo(x_1)$};
   \draw (6,1) ellipse (0.2 and 0.12);
  \draw (6-0.1,1-0.1) -- (6+0.1,1+0.1);
  \draw (6-0.1,1+0.1) -- (6+0.1,1-0.1);
   \node at (6,1.6) {$\Phi(X)$};
       \end{tikzpicture}
    \end{center}
    \vspace{-3mm}
    \caption{\addtolength{\baselineskip}{-.5mm}{The uniformization theorem provides a map from a bulk-to-boundary 2-point function $\la \Omega\li \Phi(g) {\cal O}(x_1)\li \Omega \ra$ in a Ba\~nados geometry to a 2-point function $\la 0 \li \Phi(X) {\cal O}(x_1)\li 0 \ra$ in AdS${}_3$. The bulk point $X_0$ depends on both $g_0$ and $\Omega$.
The map makes essential use of Virasoro symmetry.}}
\end{figure}
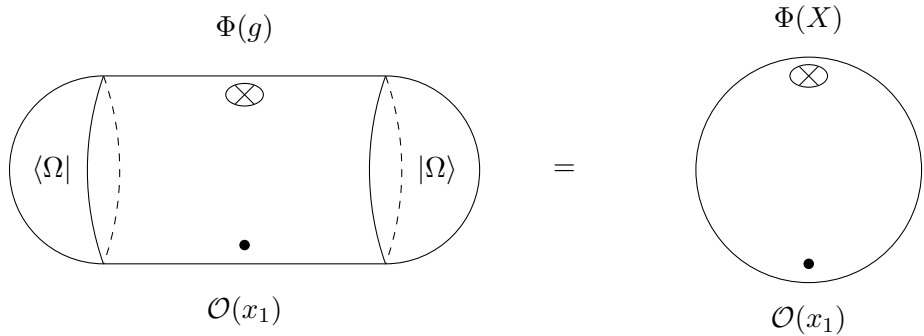

In this way, we have demonstrated that locally we can always go to the uniformizing coordinate system, in which the 2-point function (\ref{twopt}) reduces to the standard 2-point function between two $SL(2,\mathbb{R})$ invariant vacuum states 
\bea
\label{newpt}
\la \spc \Omega  \spc \ri \spc \Phi(g)\, {\cal O}(x) \spc  \li \spc \Omega \spc \ra \is 
\la \spc 0  \spc \ri \spc \Phi(X)\, {\cal O}(x) \spc  \li \spc 0 \spc \ra.
\eea
As explained before, the right-hand side is easily computed by going to the Schottky double, producing an answer that agrees with the bulk-to-boundary propagator and  locally satisfies the wave equation in AdS${}_3$.  Globally, we need to take into account the monodromy  of the uniformizing coordinate $Z$. In particular, if the state $\li \Omega\ra$ corresponds to a BTZ black hole,  the 2-point function (\ref{newpt}) matches with the bulk-to-boundary propagator in the BTZ background geometry \cite{HV}.

\subsection{Uniformizing gauge}\label{SecUniGauge}

As mentioned in the introduction, gravitational dressing is the analogue of turning a local observable, like a charged matter field, into a gauge invariant observable.
This can be done either by attaching a Wilson line or a Coulomb field. By choosing a particular gauge, like an axial or Coulomb gauge, the non-local  gauge invariant observable 
can be made to look like an ordinary local observable. We would like to find a similar gauge for our problem. We will call it the uniformizing gauge.

The above geometric definition of the gravitational dressing is indeed associated with a natural gauge choice. Once we will choose this gauge,
the dressing operator ${\mathbf 1}(X)$ is essentially set equal to unity. The gauge symmetry that we will use to fix this gauge is diffeomorphism invariance.
We can distinguish diffeomorphism that are trivial on the AdS boundary and diffeomorphism that reduce to conformal transformations on the AdS boundary,
and thereby act non-trivially on the correlation functions of the CFT.

Consider some state $| \Psi \rangle$ in the CFT. The classical expectation value of the stress energy tensor 
\bea
\la T(z)\ra  \is \la\Psi \ri T(z) \li \Psi \ra 
\eea
 is locally an analytic function of the coordinate $Z$. It transforms non-trivially under conformal transformations $
T(w) dw^2 = \bigl(\spc T(z)+ \frac c {12} \{ w, z\}\bigr) dz^2,$
with  $ \{w,z\} = \frac{w'''}{w'} - \frac 3 2 \bigl(\frac{w''}{w'}\bigr)^2$ the Schwarzian derivative. Hence the expectation value 
$\la T(z)\ra$ depends on the coordinate system.  It is a well know fact that locally we can always find a special coordinate system $\ZZ(z)$ such that
\bea
\la T(\ZZ)\ra \is 0.
\eea
We will call this the {uniformizing} coordinate system. The standard method for finding the uniformizing coordinate system is to consider the second order differential equation
\bea
\label{monodromy}
\bigl( \partial^2 - b^2\,  T(z) \bigr)\psi(z) \is 0, 
\eea
with $b$ determined by the relation $c=1+6(b+ 1/b)^2$. This is the familiar null state equation $L_{-2} - b^2 L_{-1}^2 = 0$ that forms the basis of the so-called monodromy method for computing conformal blocks. Equation (\ref{monodromy}) has two independent solutions $\psi_1(z)$ and $\psi_2(z)$. The uniformizing coordinates are then defined as the ratio of the two solutions
\bea
\ZZ(z) \is \psi_1(z)/\psi_2(z).
\eea
For a general state $| \Psi\rangle$ created, say, by acting with a collection of local CFT operators ${\cal O}(x_i)$ on the CFT vacuum, the uniformizing coordinates are multi-valued.
In passing around one of the local operators, the coordinate $\ZZ$ typically undergoes a non-trivial monodromy in the form of a M\"obius transformation  $Z \to (aZ+b)/(cZ+d).$

Now will now describe a gauge choice that circumvents this obstruction. Consider some general correlation function of $N$ bulk operators $\Phi(X_k)$ located at $N$ bulk points $X_k$. To specify our coordinate system, we first replace all bulk operators $\Phi(X_k)$ by identity operators $\mathbf{1}(X_k)$. 
The CFT correlation function $\langle {\bf 1}(X_1)\ldots {\bf 1}(X_N)\rangle$  is the chiral CFT partition function on the associated Schottky double $\widehat{\Sigma}$,
projected on the identity sector in each of the $N$ channels. We now
choose our coordinate system on $\widehat{\Sigma}$ such that
\bea
\label{gaugechoice}
\la\smpc T(\ZZ)\; {\bf 1}(X_1)\ldots {\bf 1}(X_N)\smpc \ra \is 0.
\eea
Note that this coordinate system depends on the positions $X_i$. We claim that, for CFTs at large $c$ and a sparse low energy spectrum, the above coordinate choice is globally well defined on $\widehat\Sigma$.
Indeed, since the dressing operators $\mathbf{1}(X_k)$ are all given by sums of descendents of the unit operator, the uniformizing coordinate $Z$ has a trivial monodromy 
around each operator $\mathbf{1}(X_k)$. This guarantees that, for CFTs with a gravitational dual, the correlation function $\langle {\bf 1}(X_1)\ldots {\bf 1}(X_N)\rangle$ is dominated by the identity conformal block in {\it all} channels. In the gravity theory, this means that the double cover of the bulk saddle looks like pure handle body. Hence the uniformizing coordinate $\ZZ$ defined via (\ref{gaugechoice}) has no $SL(2,\R)$ monodromies and is globally single valued. 

As we have seen in the previous section, the stress energy tensor tells us how correlation functions depend on the metric, and thereby on all geometric parameters of the correlation functions, including the position of operators. From this relationship, we conclude that with our gauge choice (\ref{gaugechoice}), the correlation function $\langle {\bf 1}(X_1)\ldots {\bf 1}(X_N)\rangle$ is independent of the $X_k$. So we~can~set
\bea\label{gaugeo}
\la {\bf 1}(X_1)\ldots {\bf 1}(X_N)\ra \is 1
\eea
In practice this means that we define the canonically normalized bulk correlation functions as the ratio 
\bea\label{normalization2pt}
\la {\Phi}(X_1)\ldots {\Phi}(X_N)\ra_{\rm bulk} \is \frac{\la {\Phi}(X_1)\ldots {\Phi}(X_N)\ra{}_{\rm cft}\!\!}{\la {\bf 1}(X_1)\ldots {\bf 1}(X_N)\ra{}_{\rm cft}\!\!}
\eea
Taking the ratio of CFT correlation functions is natural, because it eliminates the ambiguity in the overall normalization due to the conformal anomaly.\footnote{More generally, for mixed correlation functions involving $N$ bulk operators $\Phi(X_i)$ and M boundary operators ${\cal O}(x_j)$, we will be able to make the gauge choice
$
\la {\bf 1}(X_1)\ldots {\bf 1}(X_N) \, {\cal O}(x_1) \ldots {\cal O}(x_K) \ra =\la  {\cal O}(x_1) \ldots {\cal O}(x_K) \ra
$}

\section{Bulk two-point function} \label{sec2ptbulk}

In this section we will study the two point function of two bulk operators.  The classical bulk geometry corresponding to the Klein bottle has been described and analyzed in detail in \cite{MR}.  It has two fixed points, that we interpret as the location of the bulk fields. In \cite{MR} it is shown that the semi-classical partition function associated with this saddle point, computed  at one loop, exactly matches the identity conformal block 
 in the CFT.  We have given this result a new interpretation as the two-point function of two local $\Z_2$ defect operators ${\mathbf 1}(X)$ placed at the corresponding bulk points.
Following the discussion in section \ref{SecUniGauge}, we will normalize correlation functions according to \eqref{normalization2pt} which is equivalent to choosing the physical gauge such that the two point function
of two unit operators equals unity.

\subsection{Bulk to bulk propagator}\label{SecBulkBulk}

Next we turn to the bulk-to-bulk two point function.  
Adding two cross-caps to a  sphere creates a Klein bottle $K^2
=T^2/\mathbb{Z}_2$ \cite{blum}. Explicitly, we start from the torus $\tau   \to \tau + \beta,$ $\sigma \to \sigma+ 2\pi$ and make a $\mathbb{Z}_2$ identification $\tau \to -\tau$, $\sigma  \to \sigma + \pi$. The CFT partition function on the Klein bottle is then defined as $
Z = {\rm Tr}\bigl( \Omega ~ {\tilde q}^{L_0 - \frac{c}{24}} ~\bar{\tilde{q}}^{\bar{L}_0 - \frac{c}{24}} \bigr),$ where $\tilde{q} = e^{2\pi/ \beta}$ 
where $\Omega$ is a parity changing operator. Applying a modular transformation, we can rewrite this as a sum of matrix elements of the evolution operator ${q}^{L_0} \bar{q}^{\overline{L}_0}$ between cross-cap Ishibashi states $|h\rangle\! \rangle$.
Therefore the CFT partition function, after projecting onto one given primary channels, takes the form of the two point function $\lb \Phi(1,0,0)| \Phi(q,0,0)\rb$
between two local bulk states $ |\Phi(q,0,0)\rb=q^{L_0 + \overline{L}_0} | h \rb\!\rb$
at position $X_1=(1,0,0)$ and $X_2 = (q,0,0)$. We review this derivation explicitly in the appendix \ref{secKB}. 
The overlap between two cross-caps is equal to the corresponding Virasoro character 
\bea\label{vch}
\la \smpc \Phi_1\smpc \ri\smpc   \Phi_2\smpc \ra \! \is\! \chi_h(q^2) \, = \, \frac{q^{2h-\frac{c}{12}}}{\eta(q^2)}, \\[2.6mm]
 \la \smpc {\mathbf 1}_1 \smpc \ri \spc {\mathbf 1}_2 \ra \! \is \! \chi_0(q^2)\, = \,  q^{-\frac{c}{12}}\frac{1-q^2}{\eta(q^2)}\, .
\eea
These characters were computed in the Klein-bottle. Going to the plane would eliminate the background energy $-c/12$. After normalizing we see that this background-dependent factor cancels. In the physical gauge (\ref{gaugeo}), the normalized two point function between two bulk operators is obtained by taking the ratio
\bea\label{2pp}
\lb\smpc \Phi(X_1)\smpc \Phi(X_2)\smpc \rb \is \frac{Z_h}{Z_0} =\frac{\la \smpc \Phi_1\smpc \ri\smpc   \Phi_2\smpc \ra}{\la \spc {\mathbf 1}_1 \smpc \ri \spc {\mathbf 1}_2\spc \ra}\; = \; \frac{q^{2h}}{1-q^2}.
\eea
This answer should be compared with well-known result for the bulk-bulk propagator in AdS$_3$
of a field with mass squared $m_h^2 = 2h(2h-2)$
\beq\label{p1p2}
\lb \Phi(X_1) \Phi(X_2)\rb = \frac{e^{-(2h-1)D(1,2)}}{2 \sinh{D(1,2)}}.
\eeq
where $D(1,2)$ is the geodesic distance between the two points $X_1$ and $X_2$ in AdS$_3$, which in Poincar\'e coordinates is given by
\beq\label{geod}
\cosh{D(1,2)} = \frac{y_1^2 + y_2^2 + x_{12}\bar{x}_{12}}{2y_1y_2}.
\eeq
which suggests that we should make identification $q(1,2) = e^{-D(1,2)}.$
It is straightforward to verify this relation, by performing a conformal map from the annulus to the Schottky representation of the Klein bottle in terms of two circles.
This analysis in performed in the Appendix. We find that
\bea\label{qads}
q(1,2) \! \is \! \frac{y_1^2 + y_2^2 + x_{12}\bar{x}_{12} -\sqrt{( y_1^2 + y_2^2 + x_{12}\bar{x}_{12})^2-4 y_1^2y_2^2}}{2y_1y_2} \, = \, e^{-D(1,2)},
\eea
which establishes the identification between (\ref{2pp}) and (\ref{p1p2}).
The result \eqref{p1p2} can be derived equally for global cross-caps and gives the same result since the global character is $\chi^g_h(q)=\frac{q^h}{1-q}$. This coincides and gives a more geometrical proof of the two point function of two global cross-caps obtained in \cite{Takaya1}.

We end with a few short comments.
The above calculation gives further support for the proposed identification of cross-cap states with bulk operators. Not only does it reproduce the bulk-to-bulk propagator, it also illustrates the mechanism that makes the background independence of the cross-cap operators manifest.
Note that the exact match between the normalized 2-point function (\ref{2pp}) with the bulk propagator holds for any CFT with $c>1$. The special assumption that relies on large $c$ and a sparse low energy spectrum is that there exists a uniformizing coordinate choice such that $\la T(z)  {\bf 1}(X_1) {\bf 1}(X_2)\ra = 0$, which enables us to pick the gauge $\la {\bf 1}(X_1) {\bf 1}(X_2)\ra =1$.

It is known \cite{Ishibashi} that Ishibashi states  $ |\Phi(y,0,0)\rb=y^{L_0 + \overline{L}_0} | h \rb\!\rb$ with size $y=1$ have infinite norm. This has an interpretation in the Poincare patch of AdS. If we put a point in the origin of size $y$ and another at infinity with size $y$, then if we send $y\to 1$ the two bulk points would approach each other.  The divergent norm thus originates from the short distance singularity of the scalar field two point function in AdS. This also appears in the global patch as a divergence in the norm of the state $| \Phi(X)\rb$ which can be regularized with a UV cut-off \cite{Takaya2}.
 
\subsubsection{$\la \Phi \Phi T\ra$} \label{SecPhiPhiT}

Next we again turn to the stress tensor insertions. The easiest case is a single insertion. Since the stress tensor generates deformations of the complex structure of the 2D space-time, this one-point function can be obtained from taking the derivative of the partition function with respect to the moduli of the surface, which in this case is $\log q$. Let us start with the geometry of the plane with the cross-caps at $0$ and infinity and with sizes $q$ and $1$ respectively. For this particular case the one point function is 
\beq\label{Tdq}
z^2\lb T(z) \rb = \frac{q}{2} \frac{\partial}{\partial q} Z_h.
\eeq
If we take for $Z_h$ the normalized partition function (\ref{2pp}) we get the result 
\beq
\lb \Phi(1) | T(z) | \Phi(q)\rb = \frac{(h+q^2-hq^2) q^{2h}}{z^2(1-q^2)^2}.
\eeq

Again we would find the same answer if we use the global cross-cap.  

For this configuration we again see that there is not unphysical branch-cuts and the poles are at the origin and at infinity, where the operators and their $\Z_2$ images are located. We can generalize this for an arbitrary configuration in the following way.  We can find the correlation function
\beq
\lb T(z) \Phi(X_1)\Phi(X_2)\rb
\eeq  
by the same method outlined above, by performing a M\"obius transformation that maps the annulus to two arbitrary circles (see Appendix). If we relabel the original coordinate as $\zeta$ then it is given by $\zeta \to z = \frac{a~ \zeta+b}{c ~\zeta +d},$
where $a,b,c,d\in \mathbf{R}$ depend on the coordinate $(1,2)=(y_1,x_1,\bar{x}_1,y_2,x_2,\bar{x}_2)$. We will not write down the explicit dependence to avoid cluttering.  The correlation function is 
\beq\label{TPhiPhieq}
\lb T(z) \Phi(X_1)\Phi(X_2)\rb = \frac{1}{\zeta(z)^2} \left(\frac{d\zeta}{dz}\right)^2\left( 2h+ \frac{e^{-D(1,2)}}{2 \sinh{D(1,2)}}\right)\frac{e^{-(2h-1)D(1,2)}}{2 \sinh{D(1,2)}},
\eeq
where $D(1,2)$ is the geodesic distance introduced above in equation \eqref{geod}. As a function of the stress-tensor insertion position $z$ this expression has poles at the zeros and pole of the M\"obius transformation because of the $\zeta(z)^{-2}$ and $ \left(\frac{d\zeta}{dz}\right)^2$ factors, respectively. For case that the bulk operator are placed on the real line $x_i = \bar{x}_i$ it is straightforward to find an expression for the position of the poles 
\beq\label{poles}
z_\pm =\frac{x_{12}^2 + y_2^2-y_1^2 \pm \sqrt{(x_{12}^2+y_1^2+y_2^2)^2-4y_1^2y_2^2}}{2 x_{12}} .
\eeq
It is easy to check that $z_+$ is the image of $z_-$ with respect to any of the two cross-cap identifications. This forces $z_+$ to be inside one cross-cap and $z_-$ inside the other. From this we conclude that, when restricted to the region outside the cross-caps, the  correlation function is again free of branch cuts and poles.

\section{Concluding remarks} \label{Conclusions}

In this paper we have examined the correlation functions and bulk properties of CFT operators $\Phi(X)$ that create cross-cap boundary states, and compared these findings with physical properties one would expect for gravitationally dressed bulk operators. We have found that the proposed identification passes a number of non-trivial tests.
Some of these checks simply follow from the fact that at large $c$, the Virasoro cross-cap boundary state $\Phi(X)$ formally reduces to a global cross-cap boundary state
$\Phi^{\mbox{\ttiny $(0)$}}(X)$, which is known to coincide with the free HKLL bulk operator \cite{NO1, NO2, Takaya1}. Our proposal is that the 
 $1/c$ correction terms that turn the global boundary state into the Virasoro boundary state incorporate gravitational dressing. 

A central theme in our story is that bulk locality, which often is seen as an organizing principle of the bulk reconstruction program, can not be expected to be an exact property of a gravitational theory.  So instead of imposing locality by hand, we have followed a more pragmatic approach that takes direct guidance from geometric and analytic properties of the CFT. 

We have found that our geometric definition of bulk operators incorporates, through concrete CFT mechanisms, two
key physical properties: background independence (the fact that $\Phi(X)$ solves the correct bulk equation of motion for any deformed vacuum state) and micro-causality (the fact that the three point function $\la\Phi(X)\calo(x_1)\calo(x_2)\ra$ has no branch cut singularity when $x_1, x_2$ and $X$ all lie on the same geodesic). Background independence and micro-causality are both highly restrictive conditions, and our results give supporting evidence that the holographic cross-cap states  defined by (\ref{treflect}) and  (\ref{holoboots}) indeed represent gravitationally dressed bulk operators in the interacting bulk theory.

We would also like to point out that the HKLL prescription to impose microcausality, while before it was directly motivated by the expectation of bulk locality, can be now understood as a natural one from the CFT point of view when trying to solve the bootstrap \cite{Polyakov:1974gs, ElShowk:2011ag, Gopakumar:2016wkt}. 

Our proposal has two somewhat unexpected features. The first is that the gravitationally dressed identity operator is in fact non-trivial. This looks surprising, but may be inevitable. In any diffeomorphism invariant theory, singling out a bulk point is in fact a non-trivial operation, that restricts the allowed configuration space of metrics. We have argued that our proposal performs this task in a natural way, but there may be other natural methods. It would for example be very instructive to compare our approach with the recent studies of AdS bulk physics using kinematic space \cite{kinematic}, geodesic Witten diagram \cite{Hijano:2015zsa, Hijano:2015qja} and OPE blocks \cite{radon}. Note, however, that in our discussion geodesics are not fundamental. For example, there are bulk backgrounds for which there are points not connected to the boundary by
geodesics. This would be the case for a heavy operator with mass less than  $c /12$. Of course,
since the cross-cap is a well defined boundary operator, it can capture operators which do not sit
in any boundary geodesic as opossed to other proposals like OPE blocks. This is implicit in the
uniformization discussion.

A second unexpected feature of our proposal is that the three point function of a bulk operator with two chiral CFT operators, such as stress energy tensors and currents, exhibits an antipodal singularity.  This type of non-locality also appears to be a necessary feature of gravitational dressing of local bulk operators.  Since the non-locality takes the form of a pole rather than a branch cut, it may be possible to partially hide the effect of the non-local correlation by smearing the location of the bulk operators over some small region.

An essential ingredient in our calculations is the uniformization theorem. It provides an organizing structure for analyzing CFT correlation functions, by exploiting the existence of a special coordinate system $Z$ in which the CFT state locally looks like a vacuum state, in a manner that directly mirrors the property of the gravity theory, that the geometry locally always looks like AdS${}_3$. The Virasoro symmetry of the boundary states is a prerequisite. Without it, the uniformization theorem would not be in play.

Some open questions that we leave for future study are the following:

\begin{itemize}
\item We have restricted most of our discussion to the special case of AdS${}_3$/CFT${}_2$.  Even though AdS${}_3$ gravity and CFT${}_2$ are both special,
due to  the absence of local gravitons and the presence of infinite conformal symmetry, our result can give some valuable clues for the problem of bulk construction in higher dimensions. 
Indeed, a natural proposal for gravitationally dressed bulk operators in AdS${}_{d>3}$ is to extend the global constraints (\ref{fixedo}) and (\ref{fixedt})  to a geometric cross-cap constraint
\bea
\label{hight}
\bigl[\spc T_{\mu\nu}(z) - {T}'_{\mu\nu}(\tildez)\spc \bigr]\, \li\spc \calO(\xX) \ra \is  0
\eea
where ${T}'_{\mu\nu}(\tildez)= \frac{\partial \tildez^\lambda}{\partial z^\mu}  \frac{\partial \tildez^\rho}{\partial z^\nu} \, {T}_{\lambda\rho}(\tildez)$
is the stress tensor transformed via the antipodal map (\ref{globalmap}). The global conditions (\ref{fixedo}) and (\ref{fixedt}) are the global zero modes of the local conditions (\ref{hight}). Based on our results, it is worth investigating whether operators that solve this equation indeed posses the right characteristics to support a holographic interpretation as dressed bulk operators.

\item Another open direction is to consider bulk fields with spin. In \cite{NO1}, a generalization of the global  cross cap conditions was introduced, which suggests that spin may be included by modifying the local cross cap constraint to $\bigl[\spc T_{\mu\nu}(z) - {T}'_{\mu\nu}(\tildez)\spc \bigr]\, |\spc \calO \rangle = S_{\mu\nu}(z) |\spc \calO\rangle$ with $S_{\mu\nu}$ the spin operator of the bulk field.   Testing the consistency of this type of modification, however, requires more work.  

\item We expect that the different representations \cite{error} that can be given to a bulk local field can also be understood in a way similar to section \ref{uniformization}. AdS/Rindler coordinates parametrize a bulk subregion of AdS, the  Rindler wedge. An operator inside a Rindler wedge can be represented by solving the wave equation in AdS/Rindler coordinates or in global coordinates and the support of the boundary operator will be different. There are many Rindler wedges that go through a given bulk point so there seems to be different representations, given by solving the wave equation in the respective wedge. From the boundary point of view, each wedge asymptotes to the causal domain of an interval, so if we write the boundary in coordinates that only foliate this region, the bulk dual will naturally be just AdS/Rindler. This can be achieved via a conformal transformation which maps this boundary subregion to a plane at finite temperature $2\pi$. So, we expect that constructing the crosscaps in this state corresponds to a bulk local operator in the AdS/Rindler wedge. These different representations should be equivalent after uniformizing the Rindler state and crosscap. Of course, this story works because the Rindler wedge has a geometric interpretation from the boundary point of view. More generally, one would like reconstruct operators in the entanglement wedge, but for that the modular hamiltonian is needed \cite{Jafferis:2015del}. Since this object is not geometric in general, we do not expect that entanglement wedge reconstruction with crosscaps to be more illuminating.

\item Finally, a key motivation for the recent interest in the construction of local bulk operators is the hope that they may shed light on the physical nature of the black hole interior. In light of this, it is relevant to note that there exists a rather concrete parallel between this question and our reformulation of the micro-causality constraint. As we have explained in section \ref{SecIntLoc2}, micro-causality requires that the bulk-to-boundary 3-point function $\langle \calo_i | \Phi | \calo_j\rangle$ with two light single trace CFT operators remains regular in the limit $\eta\to1$. The bulk-to-boundary 3-point function $\langle \calo_i | \Phi | \calo_j\rangle$ with two heavy CFT operators, on the other hand,
coincides with the mode function in a BTZ black hole background. As seen from the explicit expression given in equation (\ref{btzmode}),  the same regularity condition at $\eta\to 1$ directly maps to the requirement of regularity at the black hole horizon. 
When the bulk field approaches the horizon, the two heavy CFT operators are anti-podal relative to the bulk point at which $\Phi$ is located. On the CFT side, this means that $\calo_i$ and $\calo_j$ approach other on the second sheet.
The behavior of the mode near the horizon  thus maps to the behavior of the three point function in this kinematic limit.\footnote{The mode (\ref{btzmode}) in fact does exhibit a branch cut at $\eta=1$, but this is a direct consequence of the fact that is has a given asymptotic energy $\omega$. For this reason, a better probe of the near horizon region would be to consider, say, a bulk-to-boundary 2-point function.} Horizon regularity would require that the OPE limit  $\eta\to 1$ is dominated by heavy intermediate states with $h > h_i + h_j$. We leave this problem for future study.

\end{itemize}
\medskip

\begin{center}
{\bf Acknowledgements}
\end{center}
\vspace{-2mm}

We thank John Cardy, Bartek Czech, William Donnelly, Monica Guica, Sam McCandlish, Daniel Jafferis, Dan Kabat, Lampros Lamprou, Gilad Lifschytz, Hirosi Ooguri,  Eric Perlmutter, James Sully, Tadashi Takayanagi and Erik Verlinde for helpful discussions. A.L. acknowledges the support of a Myhrvold-Havranek Innovative Thinking Fellowship. The work of H.V. is supported by NSF grant PHY-1314198.

\bigskip

\begin{appendix}

\section{Cross caps and crossing symmetry}

In this Appendix we consider the holographic bootstrap condition (\ref{holoboots}) from an algebraic viewpoint, motivated by the structure of rational CFT \cite{MooreSeiberg}.  The three point functions $G_{ijk}(\eta)$ are given by conformal blocks. 
Conformal blocks in RCFT satisfy known linear transformation rules under crossing, that imply the existence of a monodromy matrix $M_{pk}$ such that
\bea
G_{ijp} (1-\eta) \is \sum_k M_{pk} G_{ijk}(\eta)
\eea
where ${M}_{pk} = {\rm M}\smpc\bigl[\! \mbox{\footnotesize $\begin{array}{cc} {i}\! & \! \overline{j} \\[-2mm]{j}\! &\! \overline{ i}\end{array}$}\! \bigr]_{pk}$ denotes the familiar fusion matrix.  The matrix $M_{pk}$ squares to unity, since it represents the effect of an involution, and therefore has eigen values $\epsilon_\alpha = \pm 1$. The eigen vectors 
define special crossing symmetric CFT cross-cap states 
\bea
\label{mboot}
|\alpha \rangle \! \rangle\, = \, \sum_p \spc U_p^\alpha\spc |p\rangle\! \rangle, 
 \qquad \qquad 
 \sum_{p}\;  {M}_{kp}\,  \spc U^\alpha_{p}\! \is\! \epsilon_\alpha \spc U^\alpha_{k}
\eea
with the property that the corresponding three point functions
are $\Z_2$ invariant
\bea
\label{crossings}
G^\alpha_{ij}(1-\eta) \is \epsilon_\alpha \spc G^\alpha_{ij}(\eta) \qquad \qquad G^\alpha_{ij}(\eta) \, = \, \sum_p U^\alpha_p G_{ijp} (\eta)
\eea
up to an overall sign $\epsilon_\alpha = \pm 1$.  In RCFT, the crossing (anti-)symmetric states form a complete basis of cross cap boundary states.

It is not known whether there exists an analogue of the monodromy matric $M_{pk}$ 
for holographic CFTs. If it does, it would be some random looking infinite dimensional unitary matrix with a few other prescribed properties dictated by the modular bootstrap. Similarly, it is not known if it is possible to find a basis of crossing (anti-)symmetric states $|\alpha \rangle \! \rangle$ such that its three point functions satisfy (\ref{crossings}). The existence of such crossing symmetric states is a weaker assumption than the existence of a monodromy matrix $M_{pk}$. Suppose that there exists a basis of crossing (anti-)symmetric states $|\alpha\rangle\! \rangle$. We can than give an alternative formulation of the holographic bootstrap condition, by imposing that $| \Phi \rangle$ can be written as linear sum 
\bea
\li \Phi \ra \is \sum_\alpha \PPhi_\alpha \spc \li \alpha \ra \!\nspc \ra.
\eea
This condition automatically implies that its three point functions can be expanded in the dual channel as in (\ref{holoboots}).
Via the relations (\ref{crossings}) we then deduce that the expansion coefficients $\PPhi_p$ and $\widetilde{\PPhi}_p$ that appear in  (\ref{holoboots}) are related via
\bea
\PPhi_k \! \is \! \sum_\alpha \PPhi_\alpha\spc U^\alpha_k, \qquad \quad \widetilde\PPhi_k \, = \, \sum_\alpha \epsilon_\alpha \PPhi_\alpha \spc U_k^\alpha. 
\eea
Assuming we would know the explicit form of the $U^\alpha_p$ matrix, the first equation should fix the $\Phi_\alpha$ coefficients via the boundary condition that $|\Phi\rangle = | h \rangle \! \rangle + O(1/N )$, and that $\PPhi_p = 0$ for $h_p < h$. 
The second equation then determines the $\widetilde\PPhi_k$ coefficients.\footnote{Of course, for an RCFT (or any CFT with a known fusion matrix $M_{pk}$)  the relation between the expansion coefficients is simply expressed as $\widetilde\PPhi_p = \sum_k M_{pk} \PPhi_k$.}
It is reasonable to assume that the $\Phi_k$ describe some random distribution supported on the entire spectrum over the CFT. In particular, the coefficients $\Phi_k$ of the light single trace operators are expected to be extremely small.

\section{Conformal Ward identity with bulk fields}\label{AppWI}

The conformal Ward identity gives a linear recursion relation that expresses the effect of adding one stress energy tensor $T_{\alpha\beta}(x)$ to some given correlation function. Via AdS/CFT, we can re-interpret this identity as expressing the universal effect of adding a boundary graviton localized at $x$, created or annihilated by $T_{\alpha\beta}(x)$, to the scattering amplitude of all particles created or annihilated by the other operators in the correlation function.  This recursion relation is the AdS analogue of
Weinberg's soft graviton theorem \cite{weinberg}. 

The soft graviton theorem in asymptotically flat space-time has recently been reinterpreted as the Ward identity that expresses the symmetry of the quantum gravity S-matrix under  the BMS supertranslations, the asymptotic
symmetry group of flat space \cite{andy}. AdS${}_3$ also has an infinite asymptotic symmetry group, in the form of 2D conformal transformations. The most direct analogue of S-matrix elements are the correlation functions of local CFT operators. The soft graviton theorem for these CFT $n$-point functions is simply expressed as the conformal Ward identity. With our geometric definition of bulk operators, we can now generalize this correspondence to correlation functions with bulk operators.

\smallskip

As an example,  consider a correlation function with a single stess tensor $T_{\alpha\beta}(x)$ and $n$ bulk operators $\PPhi_k(X_k)$. In the CFT this describes the one-point function of $T_{\alpha\beta}(x)$ on a sphere with $n$ cross-caps, projected onto a given  conformal sectors for each cross-cap.
 The bulk operators $\calO_k(\xX_k)$ each transform in some specific representation $V_k$ of the global isometry group $SO(2,2) \simeq SL(2,\mathbb{R})_L \times SL(2,\mathbb{R})_R$
of AdS${}_3$. We introduce thesymmetry generators $
\hat{t}_{\nspc i\ssca} = (\hat{\lL}_{i\rm a}, \hat{\rR}_{i \rm a})$
with {\small \sc a} = 0, 1,..,6 and {\small $a$} = 1,2,3. Here $(\hat{\lL}_{i\rm a}, \hat{\rR}_{i\rm a})$ denote the generators of $SL(2,\mathbb{R})_L \times SL(2,\mathbb{R})_R$,
that act as left- and right-invariant vector fields on AdS${}_3$, viewed as group manifold.

The soft graviton theorem for bulk $n$-point functions is a generalization of the standard conformal Ward identity of $n$ local CFT operators \cite{EO}. It takes the general form
\bea
\label{softy}
\Bigl\langle T_{\alpha\beta}(x)\spc \calO_1(\xX_1\nspc) \spc ... \spc \calO_n(\xX_n\nspc) \Bigr\rangle \is \hat{T}_{\alpha\beta}(x)  \Bigl\langle \calO_1(\xX_1\nspc) \spc ... \spc \calO_n(\xX_n\nspc) \Bigr\rangle \nonumber \\[1mm] \\[-2mm]
\qquad \hat{T}_{\alpha\beta}(x) = \sum_{i,\ssca} \phi_{\alpha\beta}^{i\ssca}(x)\spc \hat{t}_{\nspc i\ssca} \hspace{-2cm} & &\nonumber
\eea
where $\phi_{\alpha\beta}^{i\ssca}(x) = 
 \phi_{\alpha\beta}^{i\ssca}(x, \xX_1, . . . , \xX_n)$ denote a suitable set of symmetric tensors, that depend on the positions and conformal dimensions of all $n$ bulk operators.
  From the CFT perspective, this relation expresses the fact that a local infinitesimal variation of 2D metric deforms the complex structure of the sphere with 
  $n$ cross-caps (or more accurately, of its Schottky double). This space of complex structures ${\cal M}_n$ is parametrized by the locations $X_i$ of the bulk operators.  The
  $\phi_{\alpha\beta}^{i\ssca}(x)$ are the quadratic differentials that represent the cotangent vectors $dX_i \in T^*{\cal M}_n$. 
 
 Each bulk operator $\PPhi_k(\xX_k)$ imposes a cross-cap identification via an $SL(2,\mathbb{R})$ transformation 
 \bea
 \xX_i \is \left({ \begin{array}{cc}\!  a_i\! &\! b_i\!\\[.5mm]\! c_i\!&\! d_i\! \end{array}}\right)\quad \leftrightarrow \quad -\frac 1 {\bar{z}_i} = \frac{a_i z +b_i}{c_i z+d_i}
 \eea
Correspondingly, the action of the left- and right $SL(2,\mathbb{R})$ symmetry generators are linearly related
\bea
\bigl(\hat{\lL}_{i\rm a} - (\xX_i){}_{\rm a}{\!\nspc}^{\, \rm b}\spc \hat{\rR}_{i\rm b}\bigr) \spc \calO_i(\xX_i)\is 0.
\eea
Note that this relation identifies $\xX_i$ as a point on AdS${}_3$, defined as the coset $SO(2,2)/SL(2,\mathbb{R})$. The quadratic differentials are required to be covariant under all the cross-cap identifications
\bea
\label{phireflect}
 \phi^{i \rm a}_{zz}(z) dz^2 \! \is\!(\xX_i)^{\rm a}{}_{\rm b} \spc \phi^{i \rm b}_{\zb\zb}(\zb_i) d\zb^2_i, \\[2mm]
 \phi^{j \rm a}_{zz}(z) dz^2 \spc \! \is\!  \spc \phi^{j \rm a}_{\zb\zb}(\zb_i) {d\zb_i}^2\qquad \mbox{\footnotesize $j \neq i$}.
\eea
These relations ensure that the energy-momentum tensor is invariant under all $\Z_2$ involutions
\bea
\hat{T}_{zz}(z) dz^2 \is \hat{T}_{\zb\zb}(\zb_i) {d\zb_i}^2
\eea
The quadratic differentials $\phi_{\alpha\beta}^{i\ssca}(x)$ are uniquely determined by the reflection condition (\ref{phireflect}) and by the monodromy around each cross-cap,
which in turn is fixed by the conformal dimensions $h_k$ of each bulk field.

This soft-graviton theorem can be easily generalized to an arbitrary number of stress tensor insertions. The complete set of Ward identities reveals a few important lessons. First,
by summing up correlation functions with $T(z)$ and $\bar{T}(w)$ insertions, we can in principle determine the correlation function of bulk operators in 
arbitrary deformed background geometries. The fact that the Ward identities take the above geometric form implies that the bulk operators $\Phi(X)$ satisfy a form of background independence. 
Secondly,  given that the dimension of the space of complex structures of a sphere with $n$ cross-caps precisely matches with the
number of coordinates of $n$ local bulk operators, we can use the relation (\ref{softy}) and its generalizations to give direct proof that the correlation functions of $n$ bulk operators  in any background  only  depend on $n$ coordinates $X_k$. This minimality property is a direct consequence of the uniformization theorem, and relies on Virasoro conformal symmetry.

\section{Correlators with $T\bar{T}$ insertion} \label{AppTT}

\subsection{Virasoro cross-cap case}
We will give some details about the calculation of $\lb T \bar{T} \calo \Phi\rb$, using the notation in \eqref{TTbarV}. Since there is only one cross-cap we can compute this in the projective plane. This is equivalent to adding the image of the operator $\calo(x_1,\bar{x}_1)$ and the image of the anti-holomorphic component of the stress tensor $\bar{T}(\bar{w})\to T(w')$, where $w' = - \frac{y^2}{\bar{w}-\bar{x}_2}+x_2$. Therefore this is equivalent to computing $\lb T(z) T(w') \calo(x) \calo(x')\rb$ in the chiral Schottky double, which can be obtained from the following Ward identity 
\bea
\lb T(z) T(w') \calo(x) \calo(x')\rb &=& \frac{h}{(w'-x)^2}\lb T(z) \calo(x)\calo(x')\rb + \frac{1}{w'-x} \partial_x \lb T(z)\calo(x)\calo(x')\rb\nn
&&+\frac{h}{(w'-x')^2}\lb T(z)\calo(x)\calo(x')\rb + \frac{1}{w'-x'} \partial_{x'} \lb T(z)\calo(x)\calo(x')\rb\nn
&&+\frac{2}{(w'-z)^2}\lb T(z)\calo(x)\calo(x')\rb + \frac{1}{w'-z} \partial_{z} \lb T(z)\calo(x)\calo(x')\rb\nn
&&+\frac{c/2}{(w'-z)^4} \lb \calo(x) \calo(x')\rb.
\ea
This gives the result 
\bea
\lb T(z) T(w') \calo(x) \calo(x')\rb &=&   \frac{h^2}{(x-x')^{2h-4} (w'-x)^2(w'-x')^2(z-x)^2(z-x')^2}\nn
&&+\frac{2h}{(x-x')^{2h-2} (w'-x)(w'-x')(z-x)(z-x')(w'-z)^2}\nn
&&+\frac{c/2}{(x-x')^{2h} (w'-z)^4}.
\eea
Using the cross-cap identification to fix the position of the images and adding the proper Jacobians we can get this four-point function for the Virasoro cross-cap in \eqref{TTbarV}. 

The previous calculation relied on the conformal Ward identities. For the special configuration $\lb h | T(z) \bar{T}(\bar{w}) | \Phi \rb$ this can be derived in another way that illustrates the role of the mirroring property of the cross-cap. Expand $T$ and $\bar{T}$ in a Laurent series with $L_n$ and $\bar{L}_n$. Then use the cross-cap mirroring property that $\bar{L}_n =(-1)^n L_{-n}$ when acting on a Virasoro Ishibashi state. After this one can use the Virasoro algebra to get the result \eqref{ttbv}. In the following section we will do this for the global cross-cap. 

\subsection{Global cross-cap case}

 To make the discussion more explicit we will compute the correlation function using the global Virasoro algebra for a particular configuration $\lb h | T(z)\bar{T}(\bar{w})  | \Phi^{\mbox{\ttiny \sc (0)}}\rb$. After performing a mode expansion of the stress tensors and using the explicit definition of the global cross-cap as a sum of normalized descendants we get
\bea
\lb h | T(z)\bar{T}(\bar{w})  | \Phi^{\mbox{\ttiny \sc (0)}}\rb &=&\sum_{n,m} \lb h | L_n \bar{L}_{m} | \Phi^{\mbox{\ttiny \sc (0)}}\rb z^{-2-n} \bar{w}^{-2-m} \nn
&=&\sum_{k,n} \frac{(-1)^k}{N_k} \lb h | L_n  | k\rb_R \lb h | \bar{L}_{n}|k\rb_L  (z\bar{w})^{-2-n},
\ea
where $N_k=\frac{k!\Gamma(2h+k)}{\Gamma(2h)}$ is the normalization of the global descendant $|k\rb=(L_{-1})^k|h\rb$. In the second line we used the fact that only terms with $n=m$ contribute. Moreover, we can notice that $L_n|k\rb \sim |k-n\rb$. Therefore we only need the case to sum over $k=n$ and we obtain 
\beq
\lb h | T(z)\bar{T}(\bar{w})  | \Phi^{\mbox{\ttiny \sc (0)}}\rb = \sum_{n} \frac{(-1)^n}{N_n} \lb h | L_n L_{-1}^n|h\rb^2 (z\bar{w})^{-2-n}.
\eeq
The inner product is given by $\lb h | L_n L_{-1}^n|h\rb = h (n+1)!$. This is easy to see since it is equal to $\lb h | [..[L_n,L_{-1}],...,L_{-1}]|h\rb$ and given that $[L_n,L_{-1}]=(n+1)L_{n-1}$ the identity can be shown by induction. This gives the result as a Taylor series
\beq
\lb h | T(z)\bar{T}(\bar{w})  | \Phi^{\mbox{\ttiny \sc (0)}}\rb = \sum_{n} \frac{\Gamma(2h)}{n!\Gamma(2h+n)} [h (n+1)!]^2 (-z\tilde{w})^{-2-n}.
\eeq
Finally, looking at the definition of Gauss hypergeometric function it is easy to check that
\beq 
\lb h | T(z)\bar{T}(\bar{w})  | \Phi^{\mbox{\ttiny \sc (0)}}\rb= \frac{h^2}{(-z\bar{w} )^2} ~_2F_1\left(2,2,2h;-\frac{1}{\bar{w} z} \right).
\eeq
Of course this just recovers the global conformal block mentioned in the discussion before \eqref{TTbarG}.

\section{The Klein bottle}\label{secKB}

In this appendix we want to provide some details about the way we computed correlation functions with two cross-caps Ishibashi states. Since it will be useful below we will start by reminding the reader the explicit solution of the Ishibashi conditions \cite{Ishibashi} for both a cross-cap and a circular boundary in the origin respectively
\bea\label{Ishisol}
|\Phi_h(y)\rb_\otimes &=&\sum_{N,i} (-1)^{N} y^{2h+2N} |h;N,i\rb_R |h;N,i\rb_L ,~~~z\to z^\star = -\frac{y^2}{\bar{z}}\label{xcapishi}\\
|\Phi_h(r)\rb_{\odot} &=&\sum_{N,i} r^{2h+2N} |h;N,i\rb_R |h;N,i\rb_L, ~~~~~~~~~~~z\to \tildez=\frac{r^2}{\bar{z}}\label{circishi}
\ea
The state $|h;N,i\rb$ is a orthonormal basis for the descendants of $h$ of level $N$, labeled by the index $i$. To the right we wrote the geometric identification on the plane to which each state corresponds to. From these expressions we can see that if we compute a correlation function with circles instead of cross-caps and perform the analytic continuation $r \to i y$ to the answer we will get the cross-cap correlation function. This is the Poincar\'e patch version of the procedure presented in \cite{NO1, Takaya1} to go from a circle to a cross-cap. This analytic continuation is also reminiscent to the one that takes $AdS$ to $dS$. 

The explicit form of the Ishibashi states \eqref{Ishisol} can be used to prove that the correlation function with two cross-caps and local operators is equivalent to placing the local operators in the Klein bottle. This is a standard calculation that can be found in \cite{blum}. For example, start from 
\beq
Z=\lb \Phi_h (1) | \Phi_h (q) \rb.
\eeq
The argument uses the fact that the match between the right and left movers in the Ishibashi states is the same as the projection that one has to take in the Klein bottle because of the presence of the parity changing operators $\Omega$. This operator has to be included to implement the identification of the boundaries of the fundamental region of the Klein bottle. Because of the match between left and right movers the only moduli is real and is given by $q^2$. Then for example for a pure Ishibashi state of dimension $h$ the result gives $Z=\chi_h (q^2)$, where $\chi_h$ is the Virasoro character of the representation. 

This was done for the configuration of a cross-cap of size $q$ and $1$ at the origin and infinity. A cross-cap breaks most of the $SL(2,R)$ symmetries from the global conformal group and therefore it is not easy to get the answer for a generic cross-cap configuration. Nevertheless the identification for circles $z\to z^\star$, called conjugation, given in \eqref{circishi} is preserved by Moebius transformations \cite{alfhors}. In the case of the two point function we analyze above, we can compute the two point function for circles instead of cross-caps, namely an annulus with radius between $q$ and $1$, then use this extra symmetry to place the two circles at arbitrary positions and sizes and finally perform the analytic continuation explained above to get the answer for two cross-caps. This is possible since Moebius transformations map circles to circles. 

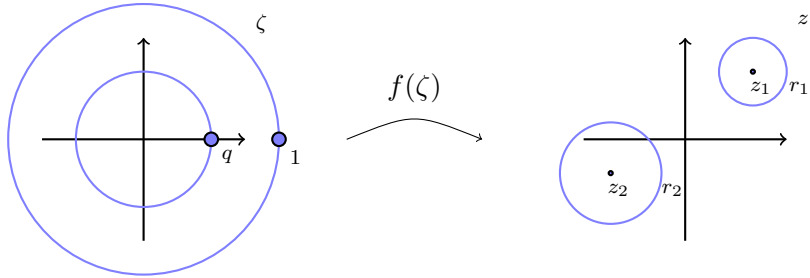
\begin{figure}[b]
\begin{center}
\begin{tikzpicture}[scale=0.9]
    \begin{scope}[thick,font=\scriptsize]   
      \draw [->] (-1.5-3,0) -- (1.5-3,0) ;
      \draw [->] (0-3,-1.5) -- (0-3,1.5);
      \path[draw,thick,blue!50] (-3,0) circle (2);
     \path[draw,thick,blue!50] (-3,0) circle (1);
     \node [below right] at (-3+1,0) {$q$};
     \node [below right] at (-3+2,0) {$1$};
     \path[draw,thick,fill=blue!50] (-3+1,0) circle (0.1);
       \path[draw,thick,fill=blue!50] (-3+2,0) circle (0.1);
      \node [below right] at (-3+1.5,2) {$\zeta$};
    \end{scope}    
   \draw[->] (0, 0) .. controls(1,0.4) .. (2, 0);
    \node [above] at (1,0.4) {$f(\zeta)$};
   \begin{scope}[thick,font=\scriptsize]   
      \draw [->] (-1.5+5,0) -- (1.5+5,0) ;
      \draw [->] (0+5,-1.5) -- (0+5,1.5);
       \path[draw,thick,blue!50] (6,1) circle (0.5);
     \path[draw,thick,blue!50] (3.9,-0.5) circle (0.75);
     \path[draw,thick,fill=blue!50] (6,1) circle (0.03);
       \path[draw,thick,fill=blue!50] (3.9,-0.5) circle (0.03);
       \node [below right] at (5.8,1) {$z_1$ \ \spc $r_1$};
     \node [below right] at (3.7,-0.5) {$z_2$ \ \ \  \spc $r_2$};
     \node [below right] at (5+1.5,2) {$z$};
    \end{scope}
    \end{tikzpicture}
    \end{center}
    \caption{\addtolength{\baselineskip}{-.5mm}{{The action of the Moebius transformation $z=f(\zeta)$ that maps the annulus between $q$ and $1$ in $\zeta$ space (left) to two circles at $z_{1,2}$ of radius $r_{1,2}$ in the $z$ space (right).}}}
\end{figure}

Knowing that the transformation exists and is unique we can find it by using the fact that it keeps cross-ratios of $4$ points fixed. In particular, we want a transformation that maps the annulus $q<|\zeta|<1$ to a circle of size $r_1$ at the position $z_1$ and another circle of size $r_2$ at a position $z_2$. For simplicity we will consider the case of $z_i$ real. Then by choosing the points $[q,-q,1,\zeta]$ in the annulus we can form the cross-ratio in the image plane giving 
\beq\label{cr}
\frac{2q(1-\zeta)}{(1-q)(\zeta+q)} = \frac{2 r_1 (z_2-r_2-z)}{(z_1+r_1-z_2+r_2)(z_1-r_1-z)}.
\eeq
From this expression we can solve for $z$ as a function of $\zeta$ which will have the usual $\zeta \to z=\frac{a\zeta+b}{c\zeta+d}$, and we can get explicit expressions for these parameters. Nevertheless, since annulus with different $q$ are not equivalent there should be a relation between $q$ and $(z_i,r_i)$. To find this we can take \eqref{cr} and use the fact that the transformation should map the point $\zeta=-1$ to $z=z_2+r_2$. Solving this equation we get 
\beq\label{q}
q= \frac{r_1^2+r_2^2 -z_{12}^2 + \sqrt{(r_1^2+r_2^2 -z_{12}^2)^2-4r_1^2r_2^2}}{2r_1r_2}.
\eeq
Now we can replace this relation in \eqref{cr} and get $z$ as a function $\zeta$, $z_i$ and $r_i$. After replacing this in the Klein bottle partition function we obtain the bulk-bulk propagator in AdS$_3$. In particular the $q$ parameter maps to \eqref{qads}, which is equal to $e^{-D(1,2)}$ with $D(1,2)$ the geodesic distance in AdS$_3$.

Of course, the fact that we can only map this annulus to circles that have a fixed $q$ parameter \eqref{q}, which after analytic continuation is the geodesic distance in AdS$_3$, is an explicit manifestation of the isometries of the bulk. 

\medskip

We can derive the correlation function between a stress tensor and two cross-caps in another way that relies on the mirroring property of the Ishibashi state instead of a Ward identity. The stress tensor as an operator satisfies $T(z)=\bar{T}(\tildez)$ when acting on the cross-cap state, where $z\to \tildez$ is the corresponding cross-cap identification
\beq\label{mirror}
T(z) | \Phi(y,x,\bar{x})\rb = \tilde{\bar{T}}\left(-\frac{y^2}{\bar{z}-\bar{x}}+x \right)| \Phi(y,x,\bar{x})\rb. 
\eeq
We can use this  to compute the overlap $\lb \Phi(y_2,0,0)| T(z) | \Phi(y_1,x,\bar{x})\rb$, which is a special case of the three-point function computed in the previous section. Apply property \eqref{mirror} to the cross-cap in the right, giving $\bar{T}$. Then apply the BPZ conjugate version of \eqref{mirror} to the cross-cap in the left, ending up with the functional equation 
\beq
\lb T(z) \rb = \frac{y_1^4y_2^4}{(x\bar{x}-\bar{x}z+y_1^2)^4}\Bigl\langle T\Bigl(- \frac{y_2^2}{\frac{y_1^2}{x-z}+\bar{x}}\Bigr) \Bigr\rangle.
\eeq 
This equation can be solved and gives 
\beq
\lb T(z) \rb = \frac{(y_1^2-y_2^2)^2}{[z(x\bar{x}+y_1^2-y_2^2)+x y_2^2 - \bar{x} z^2]^2} F(x,\bar{x},y_1,y_2),
\eeq
where $F(x,\bar{x},y_1,y_2)$ is an arbitrary function independent of $z$. From this expression we can find the poles in $z$ and recover the result \eqref{poles}. The mirroring property that fixed the poles in the correlation function is only valid for Virasoro Ishibashi states.

\end{appendix}
 \begingroup\raggedright\endgroup

\end{document}